\documentclass[conference]{IEEEtran}
\usepackage{cite}
\usepackage{amsmath,amssymb,amsfonts}
\usepackage{algorithmic}
\usepackage{graphicx}
\usepackage{textcomp}
\usepackage{xcolor}
\def\BibTeX{{\rm B\kern-.05em{\sc i\kern-.025em b}\kern-.08em
    T\kern-.1667em\lower.7ex\hbox{E}\kern-.125emX}}

\usepackage{balance}
\usepackage{kotex}
\usepackage{xspace}
\usepackage{tikz}
\usepackage{enumitem}
\usepackage[hidelinks]{hyperref}
\usepackage[ruled]{algorithm2e}
\usepackage{listings}
\usepackage{dirtree}

\newcommand{\niparagraph}[1]{\vspace{2pt}\noindent\textbf{#1}}
\newcommand*\circled[1]{\tikz[baseline=(char.base)]{\node[shape=circle,fill,inner sep=0.7pt,minimum size=1.0em] (char) {\textcolor{white}{#1}};}}
\newcommand{\bluetext}[1]{\textcolor{blue}{#1}}

\newcommand{\simname}{\textit{LLMServingSim}\xspace}
\newcommand{\simnametitle}{LLMServingSim\xspace}
\newcommand{\simnamesection}{LLMServingSim\xspace}

\makeatletter
\newcommand{\linebreakand}{%
  \end{@IEEEauthorhalign}
  \hfill\mbox{}\par
  \mbox{}\hfill\begin{@IEEEauthorhalign}
}
\makeatother

\begin{document}

\title{\simnametitle: A HW/SW Co-Simulation Infrastructure for LLM Inference Serving at Scale\\
}

\author{
\IEEEauthorblockN{Jaehong Cho}
\IEEEauthorblockA{\textit{School of Computing} \\
\textit{KAIST}\\
Daejeon, South Korea \\
\href{mailto:jhcho@casys.kaist.ac.kr}{\textcolor{blue}{jhcho@casys.kaist.ac.kr}}}
\and
\IEEEauthorblockN{Minsu Kim}
\IEEEauthorblockA{\textit{School of Computing} \\
\textit{KAIST}\\
Daejeon, South Korea \\
\href{mailto:mskim@casys.kaist.ac.kr}{\textcolor{blue}{mskim@casys.kaist.ac.kr}}}
\and
\IEEEauthorblockN{Hyunmin Choi}
\IEEEauthorblockA{\textit{School of Computing} \\
\textit{KAIST}\\
Daejeon, South Korea \\
\href{mailto:hmchoi@casys.kaist.ac.kr}{\textcolor{blue}{hmchoi@casys.kaist.ac.kr}}}
\linebreakand
\IEEEauthorblockN{Guseul Heo}
\IEEEauthorblockA{\textit{School of Computing} \\
\textit{KAIST}\\
Daejeon, South Korea \\
\href{mailto:gsheo@casys.kaist.ac.kr}{\textcolor{blue}{gsheo@casys.kaist.ac.kr}}}
\and
\IEEEauthorblockN{Jongse Park}
\IEEEauthorblockA{\textit{School of Computing} \\
\textit{KAIST}\\
Daejeon, South Korea \\
\href{mailto:jspark@casys.kaist.ac.kr}{\textcolor{blue}{jspark@casys.kaist.ac.kr}}}
}

\maketitle

\begin{abstract}
Recently, there has been an extensive research effort in building efficient large language model (LLM) inference serving systems.
These efforts not only include innovations in the algorithm and software domains but also constitute developments of various hardware acceleration techniques. 
Nevertheless, there is a lack of simulation infrastructure capable of accurately modeling versatile hardware-software behaviors in LLM serving systems without extensively extending the simulation time. 
This paper aims to develop an effective simulation tool, called \simname, to support future research in LLM serving systems.
In designing \simname, we focus on two limitations of existing simulators: (1) they lack consideration of the dynamic workload variations of LLM inference serving due to its autoregressive nature, and (2) they incur repetitive simulations without leveraging algorithmic redundancies in LLMs.
To address these limitations, \simname simulates the LLM serving in the granularity of iterations, leveraging the computation redundancies across decoder blocks and reusing the simulation results from previous iterations. 
Additionally, \simname provides a flexible framework that allows users to plug in any accelerator compiler-and-simulation stacks for exploring various system designs with heterogeneous processors.
Our experiments demonstrate that \simname produces simulation results closely following the performance behaviors of real GPU-based LLM serving system with less than 14.7\% error rate, while offering 91.5$\times$ faster simulation speed compared to existing accelerator simulators.
\end{abstract}


\section{Introduction}
Currently, there is a significant surge in efforts to exploit large language model (LLM) as a crucial component in real-world applications~\cite{naveed2024comprehensive, zhao2023survey}.
Given the prohibitively high costs associated with building on-premise infrastructure for LLM inference, the common practice is to offload LLM inference to multi-tenant ``inference serving'' systems in the cloud, exemplified by OpenAI's ChatGPT service~\cite{chatgpt}.
The massive compute and memory requirements (both bandwidth and capacity) are forcing these systems to be equipped with many AI accelerators (or NPUs).
There has been a large body of research works that aim to develop efficient hardware and software for LLM inference serving systems.
Some works target to develop customized hardware techniques for accelerating LLM inference serving~\cite{hong2022dfx, park2024lpddr}, while others focus on developing optimized system software on GPU-based scale-out systems~\cite{vllm, dao2023flashattention2, patel2023splitwise, exegpt, miao2024spotserve}.
Recently, a few pioneering works propose to take into consideration both hardware and software together for designing holistic end-to-end accelerated systems~\cite{neupims, attacc}. 
However, there is currently a lack of simulation infrastructure that allows researchers to explore their hardware-software proposals in a scale-out setting.
This limitation not only makes it difficult for architecture researchers to explore scalable accelerator solutions, but also forces system software researchers to exclusively rely on GPU-based system software in the era of specialized hardware.

This paper sets out to address this limitation and develop a LLM inference serving system simulator, called \simname, that jointly simulates the behaviors of LLM-customized accelerators and LLM inference serving system software.
\simname is built on top of an existing AI system simulator, ASTRA-sim~\cite{astrasim}, which jointly models both hardware and software for AI workloads. 
However, there are primarily two algorithmic differences, making the design principles of \simname and ASTRA-sim largely different, as described below.

\begin{itemize}[leftmargin=1.0em]
\item \textbf{Autoregressive nature of LLM generation.} ASTRA-sim focuses on distributed training, which entails millions of ``identical'' iterations of computing that simplify the simulation. On the contrary, we target LLM inference serving that involves autoregressive token generations, producing dynamically changing behaviors across different iterations, requiring independent simulation runs.
\item \textbf{Redundancies across decoder blocks in LLMs.}
Unlike ASTRA-sim that targets general models, we focus on LLMs, thus offering opportunities to customize simulations for LLM's model architecture. Modern LLMs constitute a set of decoder blocks that share common compute patterns, while the hyperparameters can vary. 
\end{itemize}

To this end, we design \simname in such a way that it prudently compromises simulation accuracy for achieving feasible simulation time, effectively bridging the so-called ``real2sim'' gap, and facilitating future research in LLM inference serving systems.
To accomplish these objectives, we exploit the following three major techniques. 

\begin{itemize}[leftmargin=1.0em]
\item \textbf{Iteration-level hardware-system simulation.}
As each iteration takes different input prompts, \simname simulates the iterations one by one temporally and aggregates the entirety of resulting statistics at the end. 
For each iteration, \simname first performs prompt scheduling that determines tasks for accelerators, then analyzes the accelerator behaviors using hardware simulator, and finally sweeps through the stages in the system pipeline to simulate overall system behaviors. 
For hardware simulation, we employ GeneSys~\cite{genesys}, an open-source end-to-end NPU simulator that comes with a full software stack. 
Note that although we use GeneSys for prototyping purposes, any NPU simulator can be integrated into \simname, as the system simulation workflow remains consistent regardless of the specific simulator employed.
The aforementioned three steps are repeated over the iterations progressively. 
\item \textbf{Compiler and simulator optimization through computation reuse.}
We notice that the hardware simulator experiences a substantial bottleneck at the compilation and hardware simulation phases. 
\simname addresses this bottleneck by optimizing implementations exploiting the redundancy of common LLM architecture and employing computation reuse techniques.
Exploiting the property that the decoder-based LLM architecture consists of repeated transformer blocks, \simname compiles just one transformer block and replicates it, significantly reducing the overall compile time required. 
In addition, we also reduce simulation time by separating simulations of attention layers from non-attention layers since it is the only computational difference between the initiation and generation phases.
\item \textbf{Operator mapping and scheduling for heterogeneous accelerator simulations.}
As modern LLM serving systems are often equipped with heterogeneous accelerators such as GPU, NPU, and PIM, accurately simulating the heterogeneous systems is an important challenge. 
While ASTRA-sim is capable of simulating heterogeneous accelerators, their operator mapping and scheduling are manually and statically determined, which would not work for LLM serving since the tasks dynamically change. 
\simname provides a flexible framework that allows users to plug in any accelerator compiler-and-simulation stacks for exploring various system designs with heterogeneous processors.
To achieve this goal, \simname comes with a ``skeleton'' interface where the simulator users can fill the system-specific operator mapping and scheduling mechanisms.
The interface connects the mapping-scheduling mechanisms with the accelerator's compiler and system simulators.
\end{itemize}

Our experiments demonstrate that the simulation results produced by \simname experience average 14.7\% error rate, showing a similar trend as in the real LLM inference serving system, vLLM~\cite{vllm}, equipped with multiple GPUs. 
Note that we observe that \simname consistently produces accurate simulation results as we vary LLM architectures, parallelization schemes, number of NPUs, and different heterogeneity. 
\simname achieves the high level of accuracy, while offering 34.7$\times$ to 491.0$\times$ faster simulation speed compared to three existing accelerator simulators: mNPUsim~\cite{mnpusim}, GeneSys~\cite{genesys}, and NeuPIMs~\cite{neupims}. 
These promising results suggest that \simname has a significant potential to be an effective system simulation tool for LLM serving system research, in hardware, software, or both. 
\simname is available at \bluetext{\url{https://github.com/casys-kaist/llmservingsim}}.

\section{Background}

\begin{figure}
    \centering
    \includegraphics[width=1.0\linewidth]{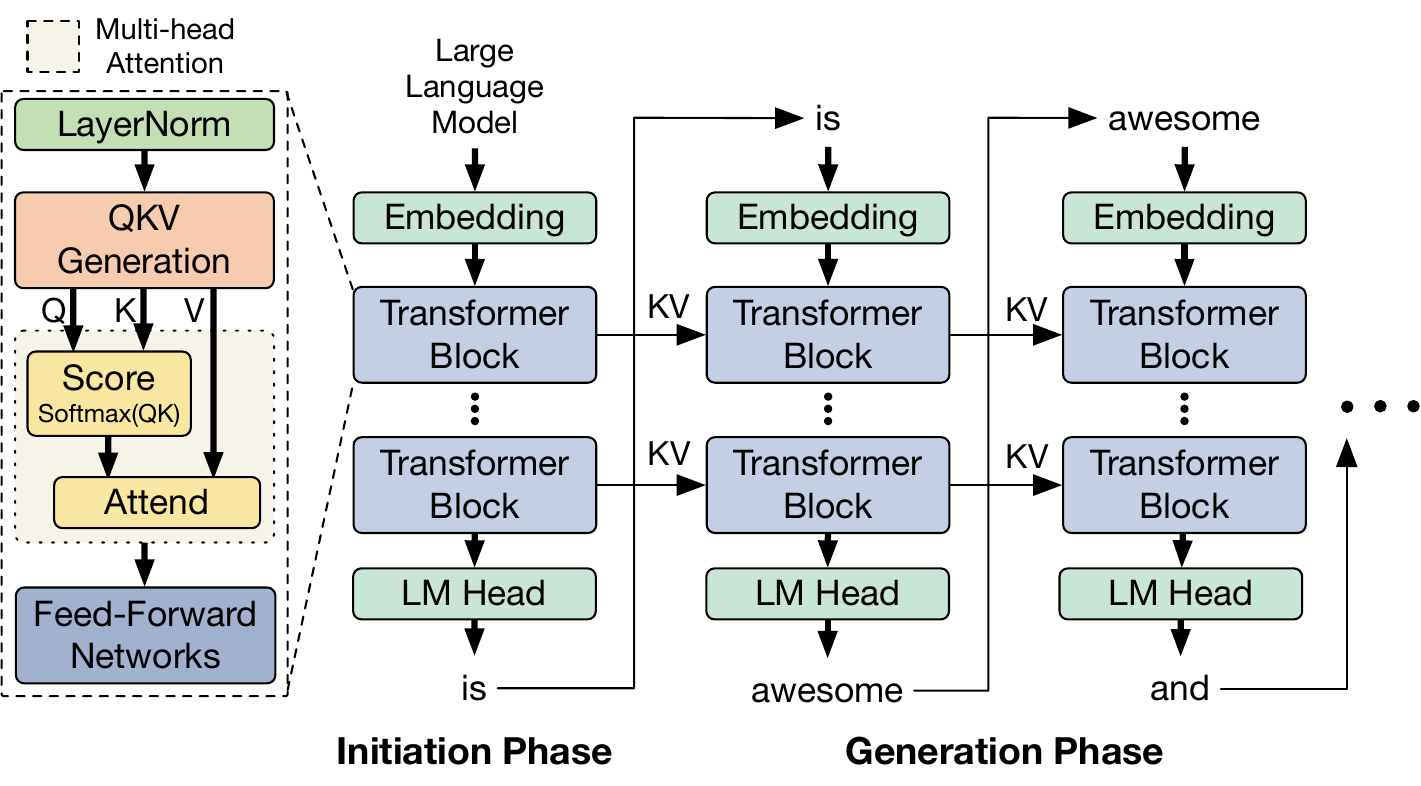}
    \caption{Architecture of large language model.}
    \label{transformer}
\end{figure}

\subsection{Characteristics of LLM Model Architecture}
\label{sec:llm-characteristics}
Most modern large language models (LLMs) employ decoder-based transformer architecture~\cite{vaswani2023attention}, as described in Figure~\ref{transformer}.
This architecture consists of its fundamental building blocks: the embedding layer, transformer blocks, and language modeling (LM) head.
Each transformer block constitutes three main components: Query-Key-Value (QKV) generation, multi-head attention, and feedforward networks.
Decoder-based transformer model operates in two distinct phases during their inference: \emph{initiation} and \emph{generation} phase.
The initiation phase begins with receiving the prompt as input and generates QKV for all input tokens.
Generated QKV values pass through subsequent multi-head attention layers and feed-forward networks.
This phase predominantly involves General Matrix Multiply (GEMM) operations, which handle the bulk of computation by processing multiple data points collectively.
Once the initiation phase is completed, the model outputs one token and transitions to the generation phase, with the generated token as the new input.
This generation phase has autoregressive characteristics where each output token is passed to the next iteration, and the generation continues sequentially.
In this phase, QKV values for newly generated tokens need to be computed, while utilizing the cached key-values of previous tokens, known as KV cache.
Consequently, this phase is characterized by General Matrix-Vector Multiply (GEMV) in \textit{Score} and \textit{Attend} operations of multi-head attention, which involve handling single vector calculations against the entire matrix of keys and values.

\subsection{Batching and Memory Management for LLM}

To minimize latency and maximize hardware utilization, LLM inference serving system often employs \emph{batching}, which involves grouping multiple requests into a single group.
However, it presents a challenge, particularly with the multi-head attention layer, which makes batching difficult.
Additionally, it faces the drawback of needing to complete all requests before proceeding to the next batch, which can lead to inefficiencies.
To tackle this challenge, Orca~\cite{orca} proposes two techniques: selective batching and iteration-level scheduling.
Selective batching allows batching in specific layers, such as QKV generation and feed-forward networks, while in multi-head attention layers, it allows a batch to be divided and allocated to multiple workers individually.
Iteration-level scheduling involves rescheduling the batch at each iteration, removing completed requests, and adding new ones.
This technique enhances hardware utilization and reduces latency by dynamically updating the batch to include only active requests, thereby streamlining the process.
Another challenge in the scale-out inference serving system is to effectively handle KV cache.
Conventional LLM serving allocates KV cache based on the maximum possible sequence length, and this results in underutilized memory spaces and limited batch sizes.
vLLM~\cite{vllm} introduces a paging scheme for memory management that functions similarly to the virtual memory of operating systems.
Managing memory on a page-by-page basis, vLLM effectively reduces memory fragmentation, enabling larger batch size and higher throughput.

\subsection{Processing-in-Memory (PIM) for LLM}

In the generation phase, LLM inferencing heavily relies on General Matrix-Vector multiplication (GEMV) operations, especially within the multi-headed attention layers.
These GEMV operations are characterized by being memory-bound with low arithmetic intensity due to the lack of matrix reuse.
Processing-in-Memory (PIM) techniques are recognized for their ability to accelerate memory-intensive operations such as GEMVs.
PIM optimizes memory-intensive tasks by reducing data movement through the placement of compute unit in each memory bank.
This approach utilizes aggregated bandwidth to read intermediate values, execute computations, and send only the results to the host system.
There has been significant research on using PIM to accelerate LLM inference.
TransPIM~\cite{transpim} has proposed a PIM-focused solution specifically for speeding up end-to-end Transformer inference.
More recently, AttAcc~\cite{attacc}, IANUS~\cite{ianus}, and NeuPIMs~\cite{neupims} have developed approaches that integrate PIM for GEMV and activation function computations, alongside compute-centric accelerators such as NPUs and GPUs, aiming to improve the overall efficiency of LLM inference.
\section{Motivation}

\begin{figure}
    \centering
    {\includegraphics[width=1\linewidth]{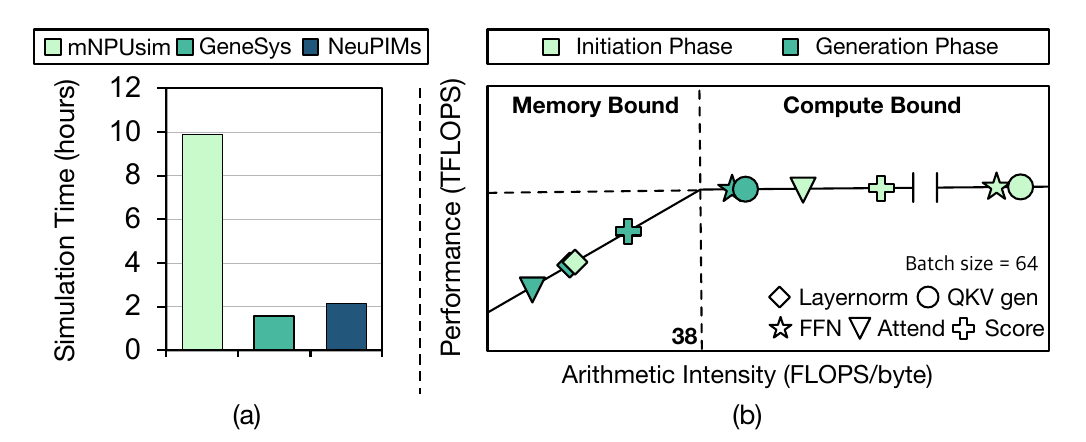}}
    \vspace{-1.5em}
    \caption{(a) Simulation time comparison between mNPUsim, GeneSys, and NeuPIMs. (b) Roofline analysis on the arithmetic intensity of LLM inference operations.}
    \label{fig:motive}
\end{figure}

\subsection{Need for LLM Serving System Simulators}
LLMs with parameters ranging from a few hundred million to several hundred billion or even trillions, require enormous computational and memory capabilities for inference.
To handle batched requests from multiple users, the scale-out serving systems often constitute hundreds of nodes, each equipped with multiple high-performance AI accelerators with high-bandwidth and high-capacity memories~\cite{park2024lpddr, nvidia-h100, neupims, attacc}.
Recently, several studies have explored solutions involving software, hardware, or both of them for such large-scale LLM serving systems~\cite{hong2022dfx, lu2020ha, ham2021elsa}.
However, the absence of an effective system-level simulator for scale-out LLM serving systems remains a major barrier for researchers and engineers who continue to explore solutions.

\subsection{Limitation of Existing AI System Simulators}
\niparagraph{ASTRA-sim.}
We are not the first one who propose to develop scale-out system simulators for AI workloads. 
ASTRA-sim~\cite{astrasim} is an effective open-source tool for simulating scale-out system for AI workloads and can be considered as an alternative to \simname. 
However, ASTRA-sim focuses on \emph{training} with repetitive yet identical iterations, which does not align well with the nature of LLM inference serving, where each iteration processes different batches with varying compositions of variable-length prompts.
While ASTRA-sim is insufficient for our purposes as it stands, we notice that it offers essential features for simulating a scale-out AI system. 
Therefore, we decided to avoid reinventing the wheel and integrate ASTRA-sim as a module within \simname to simulate a single iteration.

\niparagraph{LLM inference simulators.}
Rather disjointly, there have been recent research efforts to build LLM inference simulators~\cite{genesys, neupims, mnpusim, onnxim}, while they are not suitable for system-level simulation at scale.
The main reason is that the current simulators operate at a slow pace, rendering them insufficient for simulating large-scale LLM inference serving, which is inherently iterative in nature.
Figure~\ref{fig:motive}(a) shows the simulation time for one inference iteration of existing LLM simulators, including mNPUsim~\cite{mnpusim}, GeneSys~\cite{genesys}, and NeuPIMs~\cite{neupims}.
We observe that NPU simulators,  mNPUsim and GeneSys, take about 10 and 1.5 hours respectively, while NeuPIMs, which simulates NPU-PIM hardware, takes roughly 2 hours.
To fully process requests, multiple iterations are required until the generation phase ceases, significantly extending the entire simulation time, exceeding far beyond the simulation time reported above.
Thus, it is apparent that exploring the LLM serving system designs with these inefficient simulators is nearly infeasible.
This insight motivates us to devise a new LLM serving system simulator that enables efficient system-level exploration within feasible simulation time, while accurately evaluating the hardware-software behaviors.

\begin{figure}
    \centering
    {\includegraphics[width=0.9\linewidth]{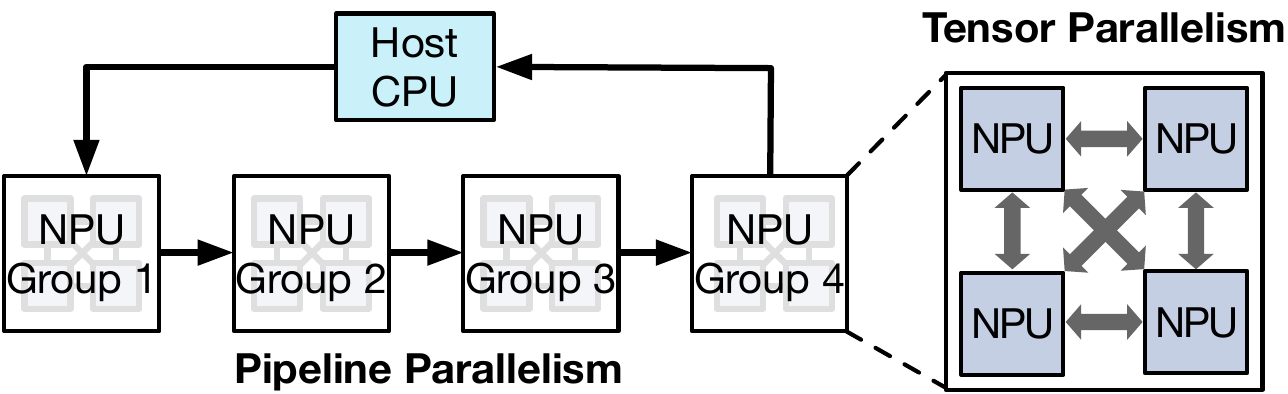}}
    \caption{Example system topology of \simname configured with hybrid parallelism, consisting of 4 pipeline parallel groups and 4 tensor parallel NPU nodes.}
    \label{topo}
    \vspace{-0.3em}
\end{figure}

\subsection{Need for Simulators with Heterogeneity}
As discussed in Section~\ref{sec:llm-characteristics}, one of the notable characteristics of LLM inference is that compute-intensive operations and memory-intensive operations are intermixed.
We analyze and compare the computation and memory usage of each operation using GPT3-7B model with NVIDIA RTX 3090 GPU.
Figure~\ref{fig:motive}(b) shows the result of roofline analysis comparing arithmetic intensity of operations during inference.
We notice that operators of multi-head attention and layer normalization have low arithmetic intensity and are bandwidth-bound.
On the contrary, operators of QKV generation and feed-forward networks have high arithmetic intensity and are compute-bound.
These two types of operators require high memory bandwidth and high compute capability, respectively.
Meanwhile, another characteristic of LLM inference is that key-value (KV) cache imposes a significant overhead on memory capacity~\cite{hooper2024kvquant, zhao2024alisa, vllm}, since keys and values are generated for every token of the entire sequence and every transformer block.
Requirements for high memory bandwidth, memory capacity, and computation power of LLM inference make it difficult to find ``one-fits-all'' solution for acceleration.
GPUs equipped with high-bandwidth memory such as NVIDIA H100~\cite{nvidia-h100} appear to be this solution, but they have small and limited scalability of capacity.
Several recent studies have proposed solutions with heterogeneous accelerators for LLM inference serving~\cite{neupims, attacc, hbm-pim, patel2023splitwise}.
A system using heterogeneous accelerators maps operators with conflicting properties to devices with different characteristics.
For instance, with inference acceleration solutions using PIM and NPU~\cite{neupims, attacc}, operators with low arithmetic intensity are mapped to PIM devices, and other operators are mapped to NPU devices.
Following this pioneering research, various combinations of heterogeneous accelerators are being explored~\cite{neupims, attacc, smart-infinity}, highlighting the need for simulator frameworks to support these efforts.
This phenomenon suggests that \simname must provide rich flexibility in system configurations, while supporting simulation of various hardware in a plugin manner.

\section{\simnamesection}

\begin{figure}
    \centering
    {\includegraphics[width=1.0\linewidth]{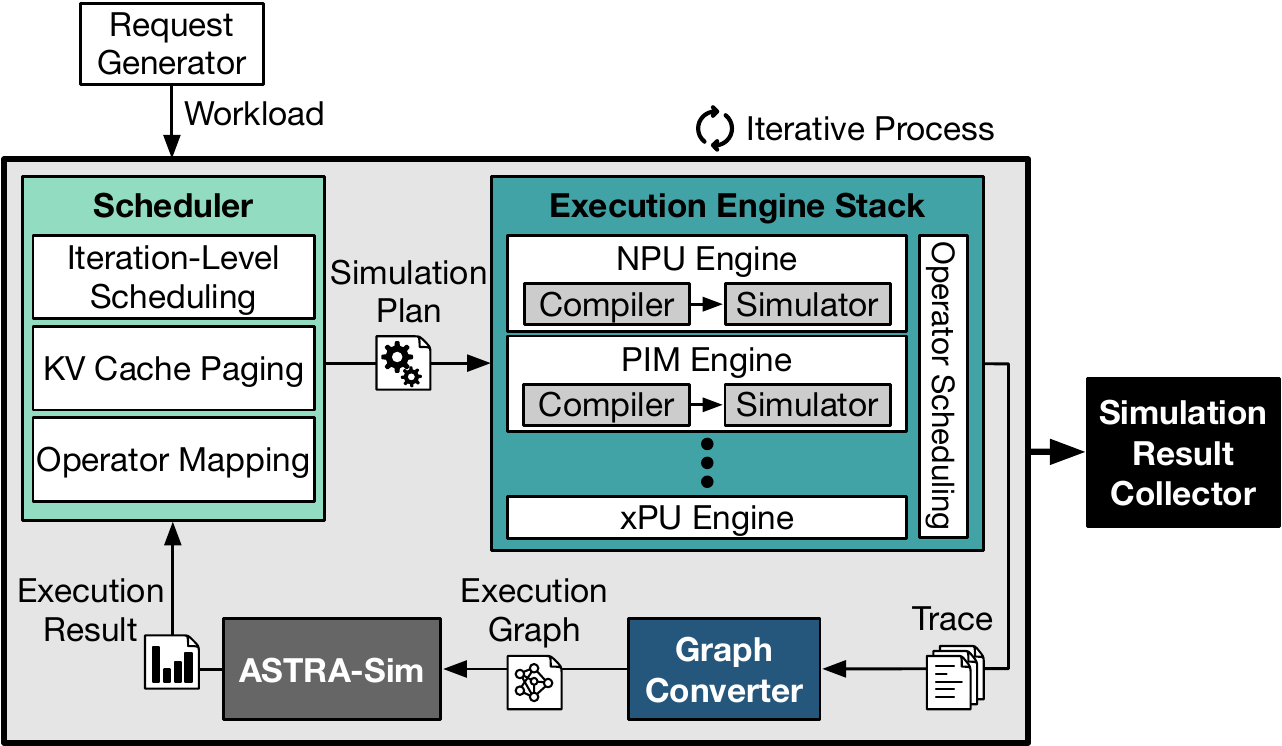}}
    \caption{Workflow of \simname.}
    \label{fig:llmsim}
\end{figure}

We design \simname, a novel system-level simulator for LLM inference workloads that jointly simulates LLM serving system software and heterogeneous hardware accelerators.
For simplicity of explanation, we provide an example where \simname simulates a distributed and heterogeneous system consisting of one host node, NPU nodes, and PIM nodes.
Figure~\ref{topo} illustrates an example \simname system topology configured to utilize hybrid parallelism with 4 NPU groups and 16 NPU nodes.
Note that \simname can be flexibly configured with various system topologies and combinations of heterogeneous accelerators.

\subsection{Simulator Design}
\label{sec:infrastructure}

\niparagraph{Overview.}
Figure~\ref{fig:llmsim} depicts the \simname workflow, which is designed to perform iteration-level simulation for distributed system with heterogeneous hardware.
\simname consists of the following components:
\begin{description}[labelindent=0.0em,nolistsep,leftmargin=1.5em]
\item[(1)] \textbf{Scheduler}
receives and organizes user requests into feasible batches based on the scheduling, KV cache management, and operator mapping strategy. It also makes the next scheduling decision based on the results of ASTRA-sim.
\item[(2)] \textbf{Execution engine stack}
compiles the model according to the batch configuration created by Scheduler, and performs hardware simulation for a single device. Each heterogeneous accelerator has distinct engine and produces distinct trace. Execution engine stack schedules operators from multiple traces and reconstructs them into a single trace.
\item[(3)] \textbf{Graph converter}
generates execution graphs using the given trace from execution engine stack, according to the configured parallelism strategy.
\item[(4)] \textbf{ASTRA-sim~\cite{astrasim}}
takes execution graph represented in Chakra graphs~\cite{chakra} as inputs, performs system simulation, and returns results back to the scheduler.
\end{description}
\niparagraph{Iteration-level scheduling.}
LLM processes input prompts autoregressively by generating one token at a time during inference. 
To efficiently process the iterations, a state-of-the-art LLM serving system, Orca~\cite{orca}, proposes iteration-level scheduling.
We employ this technique in \simname by  
designing the simulation workflow as repeated alternations of prompt batch scheduling, hardware simulation, and system simulation at the iteration level.
\simname scheduler first receives requests and compares their arrival times to the scheduler's timer to select \emph{batchable} requests.
In response to the dynamic changes in requests, the scheduler leverages execution engine stack, consisting of the engine-specific compilers and simulators, to simulate the behavior of accelerators.
In a heterogeneous environment, operators are dealt in different accelerators, so the scheduler offloads operators to each execution engine according to the mapping strategy.
Each execution engine compiles the model and simulates the hardware with specified input configurations.
After hardware simulation, the graph converter converts the simulation results to an execution graph that maps the hardware to the system.
This graph is then fed into ASTRA-sim to simulate and analyze the system behavior comprehensively.
System simulation results are fed back to the scheduler, and the scheduler's timer, which is used to assemble a new batch for the next iteration, is updated accordingly.
This cyclical interaction enables \simname to progress through iterations efficiently.

\niparagraph{Supporting for LLM parallelism strategies.}
\label{sec:parallelism}
In the context of LLM inference, parallelism that distributes the model weights and layers of substantial size is crucial for enhancing performance.
There are three major types of model parallelism: tensor parallelism, pipeline parallelism, and hybrid parallelism~\cite{megatron}.
Tensor parallelism distributes the weight matrix across multiple workers.
Pipeline parallelism assigns different layers of the model to different workers.
Hybrid parallelism combines features of both tensor and pipeline parallelism.
\simname can be configured to utilize a specific parallelism strategy by setting the number of accelerator groups according to the system's topology.
When graph converter receives the output trace from execution engine stack, it identifies configured parallelism strategy and constructs execution graph accordingly for each accelerator.
For tensor parallelism, it distributes tensors across the entire nodes and inserts \textit{ALL-REDUCE} operators to the execution graph for intermediate synchronization.
For pipeline parallelism, it allocates decoder blocks to nodes in sequence, allowing chained computation across them.
Hybrid parallelism combines both parallelism strategies by distributing tensors and layers within and across accelerator groups, respectively.
To employ selective batching, where attention layers are processed in parallel across different workers, the execution engine stack and graph converter work together.
Hardware simulator assigns unique identifiers to the attention layers and records them in the output trace.
Graph converter then assigns these attention layers to different nodes based on their identifiers.
As illustrated in Figure~\ref{topo}, each node within an accelerator group independently processes distinct inputs with different sequence lengths, parallelizing batch processing.

\begin{figure}
    \centering
    {\includegraphics[width=0.8\linewidth]{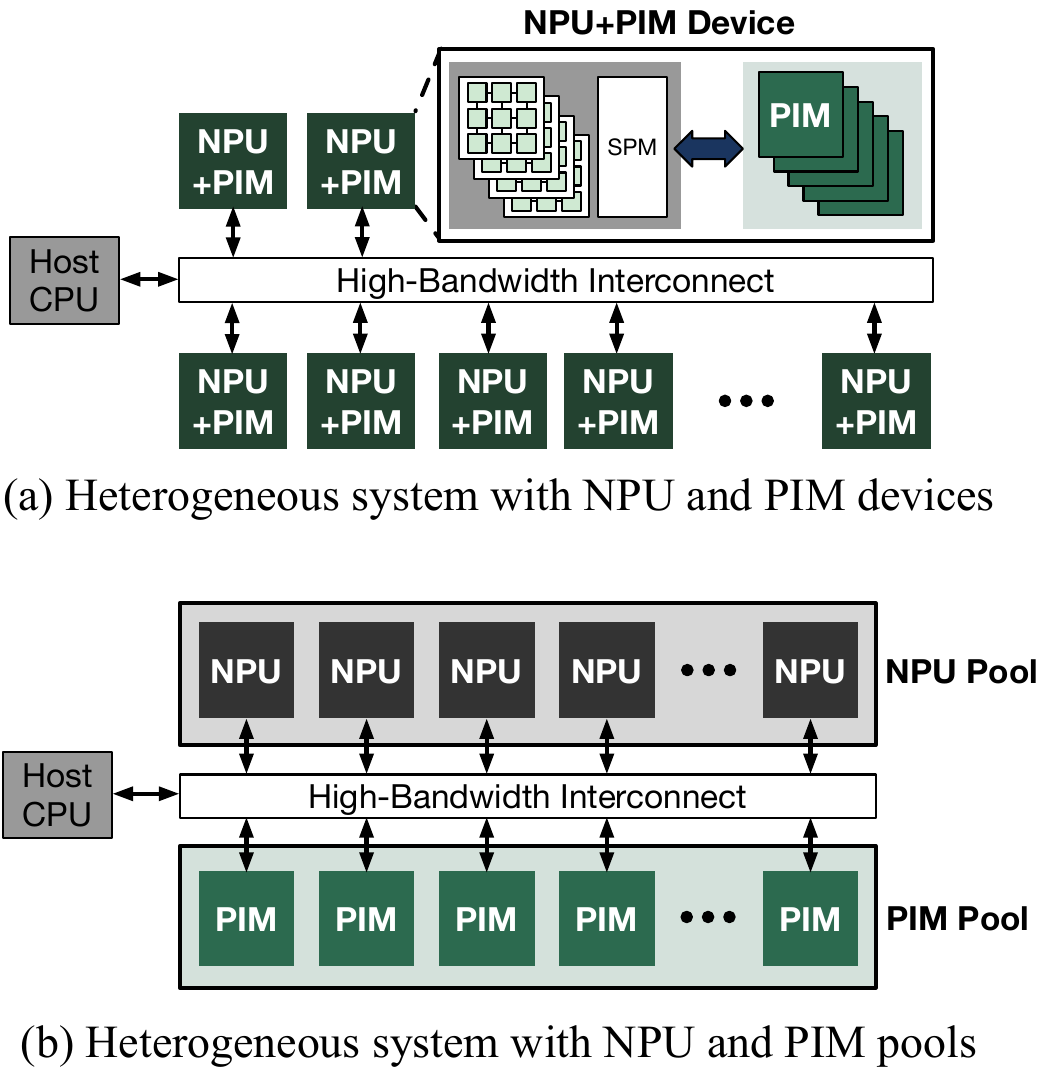}}
    \caption{Two example system topology of \simname with NPU and PIM hardware.}
    \vspace{-0.3em}
    \label{fig:npupim}
\end{figure}

\niparagraph{KV cache-aware memory modeling.}
While ASTRA-sim has a simple memory model in its implementation, it lacks some memory constraints such as capacity and memory fragmentation.
However, LLM inference is sensitive to memory capacity due to their significant memory usage of model parameters and KV cache.
\simname uses detailed memory modeling scheme with several memory constraints to reduce the gap with actual systems.
Memory model of \simname includes management of KV cache and generated tokens by incorporating demand paging technique from vLLM~\cite{vllm}.
The management of KV cache and generated tokens in \simname scheduler is intertwined with iteration-level scheduling, which conducts batch reconstruction each iteration, by checking generated tokens and KV cache size of each batch.
First, scheduler assesses the length of incoming requests to determine the required number of KV cache pages and allocates them to the local memory of accelerators accordingly to form a single batch.
After an iteration completed, the scheduler reassesses the requests.
If increased sequence length due to generated tokens requires additional page or incoming requests need to be added to the batch, new page is allocated on demand.
If there is insufficient memory capacity for new pages, the entire page for KV cache and sequence of the last added requests are evicted to host memory.
When memory availability permits, evicted KV cache blocks are reloaded from host memory for processing in subsequent batches.
The graph converter inserts operators into the execution graph for page eviction and reloading based on the decision of the scheduler.
Whenever page eviction or reloading occurs, it inserts memory store or load operators embedded with the time taken to transfer the pages between accelerator device memory and host memory into the graph.
This interaction between the scheduler and graph converter enables \simname to effectively utilize page-based memory modeling.

\subsection{Simulating Heterogeneity in LLM Serving}
\niparagraph{Heterogeneous system overview.}
\simname supports simulation of heterogeneous systems composed of two or more different types of accelerator hardware beyond homogeneous systems.
In this paper, we use example systems consisting of NPU devices for compute-bound operations and PIM devices for memory-bound operations.
While we use these particular systems as running examples, it is worthwhile to note that \simname also supports simulation with hardware accelerators other than NPU or PIM by adding new execution engine to \simname infrastructure in a plug-in manner.
As discussed in Section~\ref{sec:infrastructure}, \simname can flexibly configure the system topology.
Figure~\ref{fig:npupim} illustrates two example systems: a heterogeneous system where NPU and PIM devices are directly connected, and a heterogeneous system where there are separate pools of NPU devices and PIM devices.
For both example systems, accelerator nodes are connected to other accelerator nodes and hosts through high bandwidth interconnects such as CXL~\cite{cxl}.
Algorithm~\ref{alg:scheduler-overview} describes the overall workflow of \simname scheduler, execution engine stack, and graph converter to generate execution graph with the given request batch in the context of operator mapping and scheduling.
In the following section, we discuss how \simname's operator mapping and scheduling decisions are made depending on the system topology or computational properties of acceleration hardware.

\niparagraph{Operator mapping.}
The components that perform operator mapping could be different depending on the heterogeneous system's topology and configuration.
In \simname, the components responsible for operator mapping are execution engine, scheduler, and graph converter.
To understand how these three components interplay, we describe operator mappings in the two example systems depicted in Figure~\ref{fig:npupim}.
\niparagraph{\circled{1} Operator mapping in execution engine.}
For instance, in a heterogeneous system consisting of NPU and PIM devices, memory-bound operations such as \textit{Attend} and \textit{Score} of multi-head attention layers are mapped to the PIM module.
And, remaining compute-bound operations are mapped to the NPU module.
However, as the NPU and PIM devices are directly connected each other, they act as one node at the system-level, so there is only one execution engine in \simname.
Therefore, in the simulation of NPU-PIM system, mapping decision is done in the internal scheduler of execution engine.
\niparagraph{\circled{2} Operator mapping in scheduler.}
On the other hand, in a heterogeneous system consisting of NPU and PIM pools, operator mapping is done in two components, scheduler and graph converter, rather than execution engine.
In Line~\ref{line:operator_mapping} of Algorithm~\ref{alg:scheduler-overview}, the scheduler decides which operator will be mapped to which device by considering the characteristics of both the operators and the hardware devices, and creates simulation plans.
Then the scheduler delivers simulation plans composed of various operators mapped to the appropriate execution engine based on the operator mapping strategy.
For instance, in the simulation of a heterogeneous system consisting of NPU and PIM pools, \simname scheduler creates a simulation plan consisting of memory-bound GEMV operations and delivers it to the PIM execution engine.
Conversely, a simulation plan consisting of the remaining compute-bound operations is delivered to the NPU execution engine.
Finally, the scheduler triggers an execution engine stack consisting of compilers and simulators, each of which executes the scheduler's simulation plan to generate a trace, the input to \simname's graph converter.

\SetKwInOut{Input}{input}
\SetKwInOut{Output}{output}
\SetKwInput{KwData}{Parameter}
\SetKwComment{Comment}{\textsf{//} }{}
\SetKwComment{CommentTwo}{/* }{~*/}

\SetKwFunction{Formbatch}{\textnormal{Batch\_formatting}}
\SetKwFunction{Splitbatch}{\textnormal{Batch\_partitioning}}
\SetKwFunction{Profileops}{\textnormal{Operator\_profiling}}
\SetKwFunction{Mapops}{\textnormal{Operator\_mapping}}
\SetKwFunction{Executionengine}{\textnormal{Execution\_engine}}
\SetKwFunction{append}{\textnormal{append}}
\SetKwFunction{Batchschedule}{\textnormal{Operator\_scheduling}}

\begin{algorithm}[tb]
\footnotesize
\caption{Operator Mapping and Scheduling}
\label{alg:scheduler-overview}
\LinesNumbered

\KwIn{$
    \hspace{0.2em}{L_{req}} \ \ \ \ :\ \small\textit{A list of proceeding request information}\newline
    \hspace{0.25em}{L_{dev}} \ \ \ \ :\ \small\textit{A list of devices for mapping operators}\newline
    \hspace{0.2em}\ \ \ {Mem_{free}}:\ \small\textit{Available memory for storing KV cache}\newline
    \hspace{0.2em}\ \ \ {Time_{cur}}:\ \small\textit{Current system clock time}\newline
    \hspace{0.2em}\ \ \ {Criteria}:\ \small\textit{Key criteria to evenly partition batch}
$}
\KwOut{$\hspace{0.25em}{G_{exec}}:\ \small\textit{Execution graph for system simulation}
$}
\vspace{0.5ex}
\vspace{0.5ex}
$Batch\ =\ \Formbatch{\ensuremath{L_{req}}, \ensuremath{Mem_{free}}, \ensuremath{Time_{cur}}}$\;
$L_{sub\_batch}\ =\  \Splitbatch{\ensuremath{Batch,\  Criteria}}$\;
\label{line:batch-partitioning}
$L_{sub\_batch\_sim}\ \leftarrow\ []$\;
\vspace{0.5ex}
\ForEach{$sub\_batch$ in \ $L_{sub\_batch}$}{
    \vspace{0.25ex}
    \vspace{0.25ex}
    $L_{ops}\ =\ \Profileops{\ensuremath{sub\_batch}}$\;
    \vspace{0.25ex}
    \vspace{0.25ex}
    $L_{ops\_mapped}\ =\ \Mapops{\ensuremath{L_{ops}}, \ensuremath{L_{dev}}}$\;
    \label{line:operator_mapping}
    $L_{sim\_ops}\ \leftarrow\ []$\;
    \vspace{0.5ex}
    \ForEach{$(operator, device)$ in\ $L_{ops\_mapped}$}{
        \vspace{0.25ex}
        \vspace{0.25ex}
        $Ops_{sim}\ =\ \Executionengine{\ensuremath{operator, device}}$\;
        $L_{sim\_ops}.\append{\ensuremath{Ops_{sim}}}$\;
    }
    $L_{sub\_batch\_sim}.\append{\ensuremath{L_{sim\_ops}}}$\;
}
\vspace{0.25ex}
$Trace = \Batchschedule{\ensuremath{L_{sub\_batch\_sim}}}$\;
\label{line:operator-scheduling}
$G_{exec}= {\textnormal{Graph\_converter}}(Trace)$\;
\label{line:graph_converter}
\Return{$G_{exec}$}
\end{algorithm}

\niparagraph{\circled{3} Operator mapping in graph converter.}
\simname graph converter translates output trace into an execution graph, in Line~\ref{line:graph_converter} in Algorithm~\ref{alg:scheduler-overview}.
It embeds the type and ID of compute node where each operator will be executed into the execution graph, allowing accurate simulation based on the mapping result.
The graph converter also inserts data transfer operators to indicate the exchange of intermediate results between different accelerator pools.
In the example system shown in Figure~\ref{fig:npupim}(b), the NPU pool and PIM pool are connected through high bandwidth interconnects, so the graph converter inserts data transfer operators before and after the GEMV operator, which is processed in the PIM pool.
Operator mapping is done orthogonally to the parallelism strategy discussed in Section~\ref{sec:parallelism}, and distinct network topologies and parallelism strategies can be applied to each accelerator pool.

\niparagraph{Operator scheduling.}
Due to dependency between operators, serial execution of a batch inevitably leads to under-utilization of heterogeneous accelerators.
To overcome this limitation, \simname scheduler performs batch partitioning in Line~\ref{line:batch-partitioning} of Algorithm~\ref{alg:scheduler-overview}.
\simname scheduler splits request batch into independent sub-batches to exploit overlapping between sub-batches across heterogeneous accelerators while satisfying criteria such as fairness of computation load or memory accesses.
Subsequently, each sub-batch undergoes the operator mapping.
After operator mapping, each execution engine compiles and simulates mapped operators, and creates output trace.
These output traces include mapping and simulation information for each operator and for each hardware.
Operator scheduler of execution engine stack performs operator scheduling within a given batch by utilizing this information.
In Line~\ref{line:operator-scheduling} of Algorithm~\ref{alg:scheduler-overview}, operator scheduling decides the execution order of operators using a greedy heuristic by considering dependencies between operators and the availability of heterogeneous accelerators.
It maximizes hardware utilization of heterogeneous accelerators by allowing overlapping between operators and sub-batches.

\subsection{Techniques for Fast Simulation}
LLM typically necessitates lengthy compile and hardware simulation time. 
To solve the problem, we introduce result-reusing techniques, which reduce computation redundancy.
\niparagraph{Model redundancy reuse.}
First, we achieve significant time savings by exploiting the redundancy of common LLM architecture.
As described in Figure~\ref{transformer}, decoder-based LLM architecture consists of an embedding layer followed by repeated transformer blocks.
\simname compiles one transformer block and replicates it, largely reducing the overall compile time required.
Another optimization to reduce simulation time involves separating attention layers from non-attention layers.
The initiation phase and the generation phase differ only in attention layers, depending on the presence or absence of KV cache.
Therefore, \simname compiles and simulates the time-consuming non-attention layers just once, and subsequently, it simply swaps out the less time-intensive attention layers, cutting down on the total processing time.
\niparagraph{Computation reuse.}
Given the dynamic nature of input and output lengths in LLM inference, models typically need to be continuously compiled and simulated.
\simname adopts a strategy of reusing previously simulated results through caching.
For effective caching, it manages the non-attention and attention layers differently.
Non-attention layers take longer than other layers to be processed but can be reused frequently.
However, attention layers require more frequent compilation and simulation but take less time.
We conduct an evaluation to evaluate the impact of this caching strategy and demonstrate that our optimization technique is effective in reducing the overall simulation time.
\section{Discussion}

\subsection{Usability of \simname}
\label{sec:usability}

\niparagraph{Pluggability to 3rd-party accelerators.}
\simname's infrastructure allows for high configurability in system configuration and simulators of various hardware can be seamlessly attached via interfaces to the \simname scheduler and graph converter.
Therefore, integration with hardware simulators for various third-party accelerators other than NPU or PIM devices introduced in this paper is also possible.
Beyond acceleration hardware, it is possible to extend memory features, for instance, by adding storage capabilities like HDDs and SSDs, or incorporating computational storage nodes such as SmartSSDs.
This flexibility makes \simname a highly versatile tool for simulation and development.

\niparagraph{Compatibility with existing machine learning frameworks.}
\simname takes the ONNX\cite{onnx} model format as an input, enabling interoperability with various machine learning frameworks.
It allows users to seamlessly integrate and simulate widely-used open-source ONNX models written for frameworks such as PyTorch~\cite{torch} and TensorFlow~\cite{tensorflow}.
These models can be converted into ONNX format for use within \simname, facilitating a broad range of model experimentation and deployment scenarios.

\begin{figure*}
    \centering
    {\includegraphics[width=1.0\linewidth]{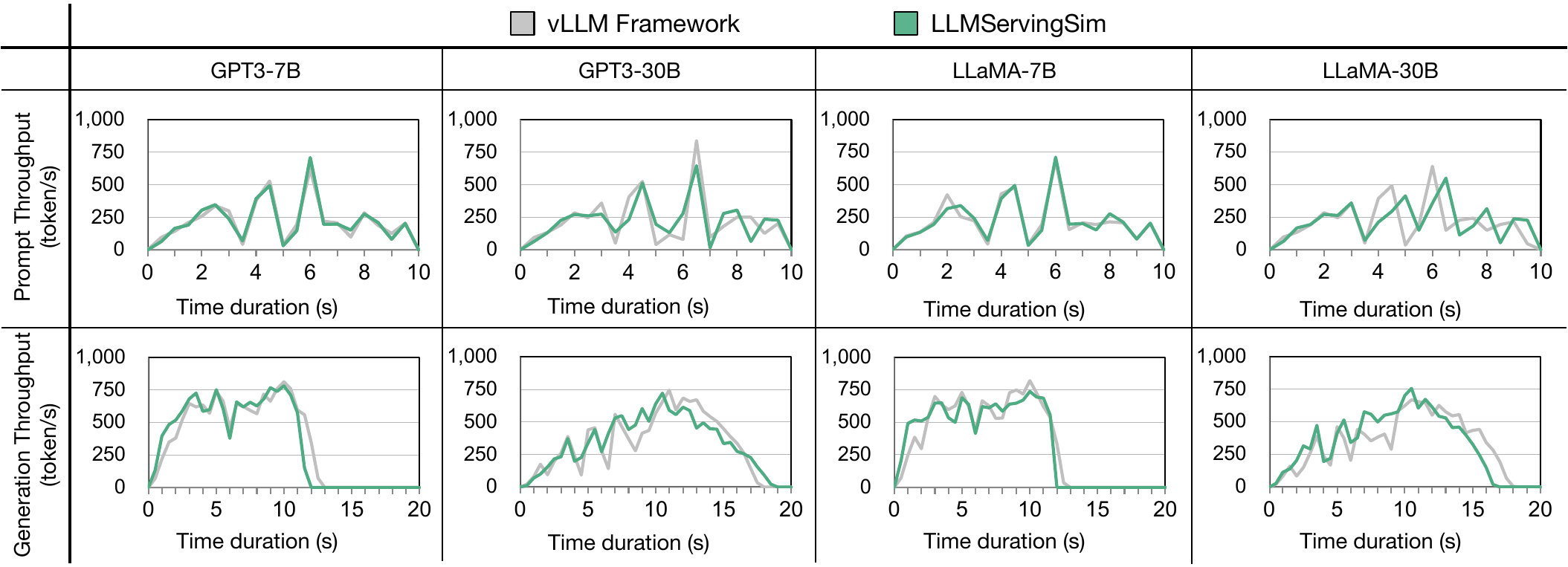}}
    \caption{Comparison of throughput over time between vLLM with GPUs and \simname using request pattern following a Poisson distribution.}
    \label{fig:tendency}
\end{figure*}
\subsection{Limitations and Future Works}
With the rapid advancement in the fields of machine learning and large language models, new variant architectures of LLM such as multi-modal~\cite{clip, minigpt4, palme, nextgpt, chameleonteam2024chameleon, peng2023kosmos2, liu2024llavanext, chen2023minigptv2}, mixture of experts (MoE)~\cite{grok, dai2024deepseekmoe, deepseekai2024deepseekv2}, and retrieval augmented generation (RAG)~\cite{chen2023benchmarking, lewis2020retrieval, shao2023enhancing, jiang2023active, xu2024retrieval, ram2023context} have been developed to address limitations of the original architecture.
Additionally, lightweight techniques such as quantization~\cite{xiao2023smoothquant, frantar-gptq, yao2022zeroquant, dettmers2024qlora, ma2024era} and pruning~\cite{frantar2023sparsegpt, ma2023llm, sun2023simple}, and fine tuning~\cite{hu2021lora, hu2023llm} techniques, offer traditional but effective solutions for optimizing LLMs.
LLMs using these architectures or technologies have different mathematical and computational characteristics compared to traditional decoder-based architectures, and ongoing studies are exploring accelerator architectures and inference systems to support these new models~\cite{patel2023splitwise, miao2024spotserve, agrawal2023sarathi, moe-lina}.
Although \simname currently focuses on traditional decoder-based LLM architectures, it can support new model variants with slight modification or even without any modification.
This is due to its inherent flexibility at the system level and its use ONNX models, which allow for flexible model construction.
For example, \simname can support MoE models by assigning each expert to one node and configuring the network topology to route to one of the expert nodes based on the inference results of the gating network.
In addition, for RAG models, it is possible to configure the system such that vector storage is simulated in a storage node, and the results retrieved from this storage are used for further inference.
Novel systems supporting these models may adopt new scheduling strategies as suggested in existing solutions~\cite{wu2023fast, duan2024muxserve, exegpt}, but we believe they can be accommodated within \simname's simulation infrastructure.
\section{Evaluation}
\label{sec:evaluation}
\begin{table}
        \centering
        \caption{\simname hardware specification.}
        \includegraphics[width=1.0\linewidth]{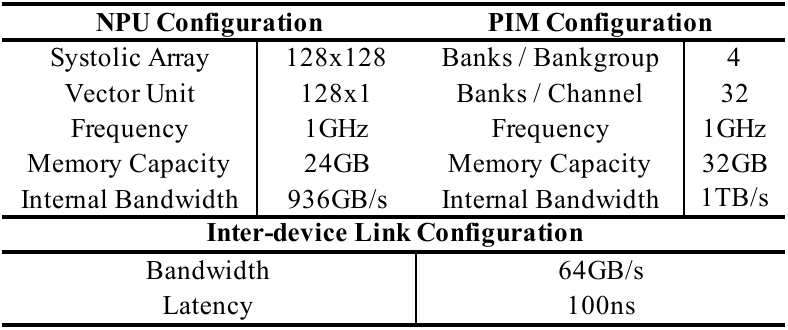}
        \label{tab:methodology}
        \vspace{-1.5em}
\end{table}

\subsection{Methodology}
\label{sec:methodology}
\niparagraph{System baselines.}
We use a homogeneous system consisting of only NPU and a heterogeneous system consisting of NPU and PIM for the validation and evaluation of \simname.
Throughout our evaluation, we use a GPU system equipped with 4 NVIDIA RTX 3090 GPUs with 24GB VRAM and Intel Xeon Gold 6326 CPU as the actual inference serving system baseline.
We use vLLM~\cite{vllm} framework as LLM inference serving system software.
This GPU system corresponds to a homogeneous system composed of multiple NPU devices, and in simulations using \simname, the performance of NPU device is set to be similar to that of the GPU.
Additionally, we use NeuPIMs~\cite{neupims}, an NPU-PIM heterogeneous LLM inference acceleration system, as one of the evaluation baselines.
\niparagraph{\simname configuration.}
\label{sec:llmconfig}
For running \simname, we use a CPU system equipped with an Intel Xeon Gold 6226R CPU with 96GB DRAM.
NPU and PIM are integrated into \simname as simulator plug-ins to the execution engine stack.
For NPU, we use PolyMath compiler~\cite{polymath} and GeneSys simulator~\cite{genesys}, and for PIM, we use an in-house PIM simulator.
Table~\ref{tab:methodology} lists the specifications of NPU and PIM hardware used throughout the evaluation.
We configure the hardware architecture of the NPU in \simname as a 128x128 systolic array with a clock speed of 1GHz to achieve similar performance to the GPU baseline, NVIDIA RTX 3090 GPU.
We set inter-device link bandwidth and latency is set to be equivalent to PCIe 4.0 $\times$16 bandwidth at 64GB/s and latency of 100ns, respectively.
In evaluation using NPU-PIM heterogeneous system, we use the same PIM hardware specification as used in NeuPIMs.
\niparagraph{Simulator baselines.}
We also compare the simulation time of \simname with other hardware simulators that support LLM inference.
We use mNPUsim~\cite{mnpusim} and GeneSys~\cite{genesys} for NPU simulation, and NeuPIMs~\cite{neupims} for NPU-PIM heterogeneous accelerator simulation.
As these baselines lack features for LLM inference serving, the simulation time for one iteration is used for comparison.

\subsection{Simulator Validation}
We evaluate the simulation accuracy of \simname against the real LLM serving systems with homogeneous or heterogeneous accelerators.
\niparagraph{NPU homogeneous system.}
Figure~\ref{fig:tendency} shows the fluctuation in throughput using a dynamic request pattern for GPT-3~\cite{gpt3} and LLaMA~\cite{llama} models, with parameter size 7B and 30B.
We synthesize request arrival patterns using Poisson distribution by sampling them from ShareGPT~\cite{sharegpt}.
We set the tensor parallelism degree as 1 and 4 depending on the model size.

In the throughput trend of initiation phase, as shown in the upper row of Figure~\ref{fig:tendency}, we observe a high degree of similarity in the prompt throughput trends between \simname and GPU-based vLLM system.
Specifically, throughput of initiation phase is influenced not only by the scheduling decision to form a request batch but also by the system's capability to accommodate the incoming requests' KV cache in memory.
Therefore, these trend results demonstrate that the iteration-level prompt scheduling and detailed memory modeling of \simname closely mirrors the behavior observed in homogeneous GPU-based system.
The lower row of Figure~\ref{fig:tendency} depicts the throughput trend in generation phase.
We observe that \simname also follows the generation throughput trend of the vLLM baseline system.
However, unlike the trend observed in the intitiation phase, there are some performance discrepancies, which can be attributed to several factors.
First, it is challenging to configure NPU architecture to precisely match the performance of the GPU.
Additionally, the degree of kernel operation optimization varies between GPU-based system and \simname.
While GPU systems often employ kernel optimization techniques such as FlashAttention~\cite{dao2023flashattention2}, the absence of such kernel optimization in \simname leads to throughput differences, especially under request-intensive conditions.
Despite the deviations, the overall throughput trend of \simname resembles that of the GPU-based vLLM, confirming that \simname can effectively simulate LLM serving systems.
\begin{figure}
    \centering
    {\includegraphics[width=0.9\linewidth]{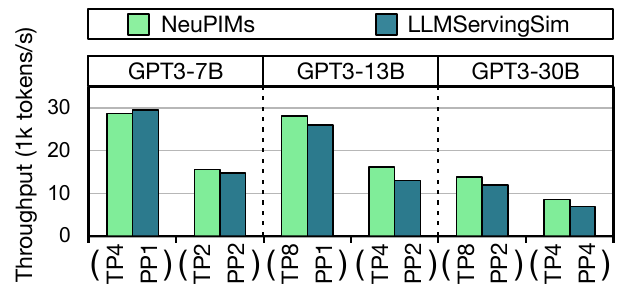}}
    \caption{Throughput comparison of \simname and NeuPIMs.}
    \label{fig:pimvalid}
\end{figure}
\niparagraph{NPU-PIM heterogeneous system.}
Figure~\ref{fig:pimvalid} shows the throughput comparison results between \simname and NPU-PIM heterogeneous accelerator, NeuPIMs~\cite{neupims}.
For the workload, we sample requests from Alpaca\cite{alpaca} and use 256 requests for each experiment.
We attach in-house PIM simulator to \simname in order to configure a heterogeneous NPU-PIM system.
We set the number of NPU and PIM devices depending on the model size, and use various parallelization schemes.
Overall, \simname shows lower throughput than NeuPIMs, since \simname focuses on implementing detailed systems features such as inter-device link and synchronization at the system level.
However, despite these features, the performance trend shown by \simname is similar to that of the NPU-PIM heterogeneous accelerator system.
Comparing using various models and parallelism schemes, both \simname and NeuPIMs demonstrate similar throughput, with error margins below 20\% and a geometric mean error rate of 8.88\%.
This indicates that \simname can effectively simulate LLM serving system with heterogeneous accelerator under varying configurations.

\subsection{Simulation Time Speedup}
\begin{figure}[t]
    \centering
    {\includegraphics[width=1.0\linewidth]{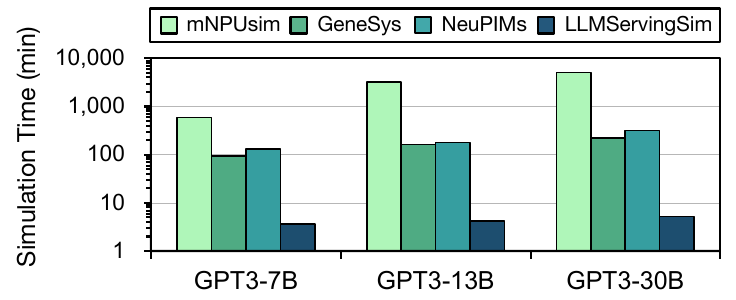}}
    \caption{Simulation time comparison of three LLM simulators mNPUsim, GeneSys, NeuPIMs with \simname.}
    \label{fig:simcomp}
\end{figure}

Figure~\ref{fig:simcomp} compares the simulation time of various LLM simulators including mNPUsim~\cite{mnpusim}, GeneSys~\cite{genesys}, NeuPIMs~\cite{neupims}, and our simulator \simname.
We measure the simulation time for one iteration of processing inputs with a batch size of 32 and a sequence length of 512 across all simulator baselines and \simname.
Also, we use GPT3 model with parameter sizes ranging from 7B to 30B.
Throughout this experiment, mNPUsim shows the longest simulation time compared to the baseline simulators and \simname.
Following mNPUsim, NeuPIMs, and GeneSys, \simname shows the fastest simulation times.
\simname show an average speedup of 490.98$\times$ over mNPUsim, 34.71$\times$ over GeneSys, and 44.97$\times$ over NeuPIMs. 
Note that, in this experiment, we assume that the initial requests have just arrived, removing the opportunities for computation reuse optimization.
Exploiting model redundancy reuse optimization has significantly boosted our simulator by skipping repeated transformer blocks.
These results illustrate \simname's superior efficiency and capability in handling large-size LLMs.

\begin{figure}[t]
    \centering
    {\includegraphics[width=1.0\linewidth]{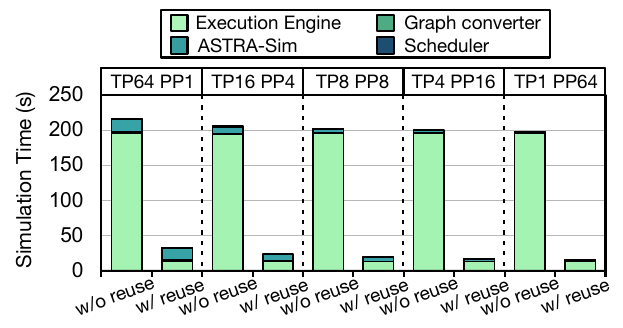}}
    \caption{Breakdown of \simname simulation time with and without computation reuse using varying parallelism strategies.}
    \label{fig:runtime}
\end{figure}

\subsection{Simulation Time Reduction}

Figure~\ref{fig:runtime} shows the simulation time and its breakdown to each component of \simname with various system configurations.
In this measurement, we use GPT-3 30B model and measure the simulation time to complete one iteration with batch size of 64 and sequence length of 1024 input.
As depicted in the graph, the running time of \simname varies significantly depending on whether reuse optimization was utilized or not.
Without reuse, running time ranges from 198.0 to 215.7 seconds, but when the optimization is enabled, it ranges from 16.3 to 33.6 seconds, demonstrating a substantial speedup of 6.4$\times$ to 12.2$\times$.
Computation reuse optimization eliminates the need to rerun the compilers and simulators in execution engine stack for each iteration.
This highlights the significant performance benefits of computation reuse optimization applied to \simname.

Figure~\ref{fig:runtime} also compares the total simulator running time using five parallelism strategies.
Execution time of ASTRA-sim for system-level simulation is the longest when using tensor parallelism solely, as it requires more synchronization operations than other parallelism strategies.
As the number of tensor parallel nodes decreases, the total simulation time also decreases, and the shortest simulation time is achieved when only pipeline parallelism strategy is used.
While there is a variance in simulation time among \simname's parallelism strategies, the difference is minimal.
This allows for the simulation of various system configurations within a feasible time, facilitating hardware and software exploration for LLM inference serving system.

\begin{figure}[t]
    \centering
    {\includegraphics[width=0.9\linewidth]{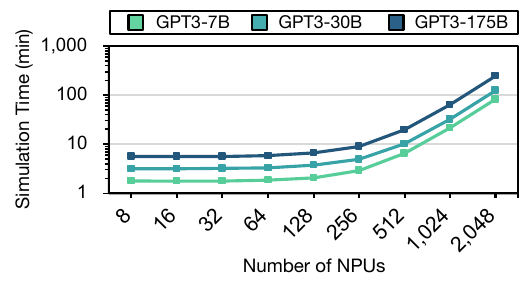}}
    \vspace{-0.5em}
    \caption{Simulation time of \simname while sweeping the number of NPU configured for tensor parallelism.}
    \label{fig:simnpu}
\end{figure}

\subsection{Simulation Time Scalability}
Figure~\ref{fig:simnpu} depicts simulation times as we sweep the number of NPUs, ranging from 8 to 2048, with the system configured to use tensor parallelism.
Additionally, we compare the simulation time using various GPT-3 models with parameter sizes including 7B, 30B, and 175B.
We use inputs with a batch size of 64 and a sequence length of 1024, and measure simulation time of one iteration.
To isolate the effect of model size and system configuration on simulation time for scalability analysis, we do not use computation reuse optimization by assuming there is no cached results for the input.
Figure~\ref{fig:simnpu} shows that the trend of simulation time tends to be proportional to the number of NPUs.
As the system is configured to use only tensor parallelism, all nodes use the same compilation and simulation results.
However, as the number of NPUs increases, system becomes more complex, requiring \simname longer time to coordinate and simulate each component.
Thus, the increased execution time is primarily due to system-level coordination and simulation of ASTRA-Sim and graph converter.
Even when scaling up the system to a vast extent using GPT-3 175B and 2048 NPUs, \simname takes 4.13 hours to simulate, outperforming other LLM simulators.
This result demonstrates the scalability of our simulator in effectively managing extensive computational loads, even at scale.

\section{Related Work}

\niparagraph{NPU simulators.}
Several simulators have been proposed to accurately model NPU behavior for ML workloads.
These simulators can either focus on a single core~\cite{samajdar2019scalesim, gem5_aladdin} or model interactions between cores in a multi-core NPU~\cite{onnxim, gemmini-dac, mnpusim}.
However, current simulators only consider the behavior of individual NPU chips and don't model systems with multiple interconnected NPU chips.

\niparagraph{Non-NPU simulators.}
With the emergence of research~\cite{ham2020a3, ham2021elsa, spatten, gobo, edgebert} on accelerating Attention operations in Transformer~\cite{vaswani2023attention} and PNM (Processing near Memory) / PIM (Processing in Memory) research~\cite{upmem-hpca24, attacc, neupims, ianus, sait-pimsim, park2024lpddr} on accelerating memory-bound operations in LLMs, there is a lot of research on measuring performance with simulation.
However, these studies have only modeled the performance of a single accelerator and lack simulations of the system.

\niparagraph{ML system simulators.}
In the realm of distributed system simulators, tools have been developed to cater to a range of needs from general-purpose workload simulators~\cite{simtool, dist-gem5, COSSIM} to those specifically designed for neural networks~\cite{dts, astrasim, shen2023mars}.
Recently, a specialized simulator tailored for LLM training~\cite{bang2023vtrain} has emerged.

\niparagraph{LLM inference serving simulators.}
Recently, reflecting the growing interest in LLM inference, a variety of simulators have been introduced~\cite{splitwise,distserve,vidur}.
However, these simulators are GPU-based and perform approximate simulation through methods like ML prediction, mathematical modeling, and using latency database instead of cycle-accurate simulations.
%

\textit{
To overcome the limitations of these existing studies, we propose \simname. 
Any hardware simulator that supports LLM operators can be integrated into our system simulator as an execution engine to perform system simulation.
\simname is system simulator that supports multi-device and heterogeneous system configurations, rather than simulating a single hardware device.
In addition, \simname successfully tackles them by exploiting techniques including iteration-level scheduling~\cite{orca}, KV cache paging~\cite{vllm}, and the interaction between hardware and system simulators.
}

\section{Conclusion}

The absence of system simulator for LLM inference serving presents challenges for researchers in system or hardware architecture exploration.
In this paper, we address these challenges by introducing \simname, a fast and accurate hardware-software co-simulation infrastructure for LLM inference serving systems, leveraging unique algorithmic characteristics of LLM serving. 
We believe that simulation tools for scale-out and heterogeneous LLM serving systems are crucial for accelerating research progress in this field, and \simname successfully makes a significant initial contribution towards meeting these needs.

\section*{Acknowledgments}
This work was supported by the National Research Foundation of Korea (NRF) (No.RS-2024-00342148), Institute of Information \& communications Technology Planning \& Evaluation (IITP) (No.RS-2024-00459797, No.RS-2024-00396013, No.2022-0-01037), and the Graduate School of Artificial Intelligence Semiconductor (No.RS-2023-00256472), funded by the Korea government (MSIT). This work was also partly supported by HyperAccel. 

\bibliographystyle{IEEEtranS}
\balance
\bibliography{IISWC24.camera_ready/paper}

\begin{thebibliography}{10}
\providecommand{\url}[1]{#1}
\csname url@samestyle\endcsname
\providecommand{\newblock}{\relax}
\providecommand{\bibinfo}[2]{#2}
\providecommand{\BIBentrySTDinterwordspacing}{\spaceskip=0pt\relax}
\providecommand{\BIBentryALTinterwordstretchfactor}{4}
\providecommand{\BIBentryALTinterwordspacing}{\spaceskip=\fontdimen2\font plus
\BIBentryALTinterwordstretchfactor\fontdimen3\font minus \fontdimen4\font\relax}
\providecommand{\BIBforeignlanguage}[2]{{%
\expandafter\ifx\csname l@#1\endcsname\relax
\typeout{** WARNING: IEEEtranS.bst: No hyphenation pattern has been}%
\typeout{** loaded for the language `#1'. Using the pattern for}%
\typeout{** the default language instead.}%
\else
\language=\csname l@#1\endcsname
\fi
#2}}
\providecommand{\BIBdecl}{\relax}
\BIBdecl

\bibitem{tensorflow}
M.~Abadi, P.~Barham, J.~Chen, Z.~Chen, A.~Davis, J.~Dean, M.~Devin, S.~Ghemawat, G.~Irving, M.~Isard, M.~Kudlur, J.~Levenberg, R.~Monga, S.~Moore, D.~G. Murray, B.~Steiner, P.~Tucker, V.~Vasudevan, P.~Warden, M.~Wicke, Y.~Yu, and X.~Zheng, ``{TensorFlow: A system for large-scale machine learning},'' \emph{arXiv preprint arXiv:1605.08695}, 2016.

\bibitem{vidur}
\BIBentryALTinterwordspacing
A.~Agrawal, N.~Kedia, J.~Mohan, A.~Panwar, N.~Kwatra, B.~Gulavani, R.~Ramjee, and A.~Tumanov, ``{Vidur: A Large-Scale Simulation Framework For LLM Inference},'' 2024. [Online]. Available: \url{https://arxiv.org/abs/2405.05465}
\BIBentrySTDinterwordspacing

\bibitem{agrawal2023sarathi}
A.~Agrawal, A.~Panwar, J.~Mohan, N.~Kwatra, B.~S. Gulavani, and R.~Ramjee, ``{SARATHI: Efficient LLM Inference by Piggybacking Decodes with Chunked Prefills},'' 2023.

\bibitem{bang2023vtrain}
J.~Bang, Y.~Choi, M.~Kim, Y.~Kim, and M.~Rhu, ``{vTrain: A Simulation Framework for Evaluating Cost-effective and Compute-optimal Large Language Model Training},'' \emph{arXiv preprint arXiv:2312.12391}, 2023.

\bibitem{gpt3}
T.~B. Brown, B.~Mann, N.~Ryder, M.~Subbiah, J.~Kaplan, P.~Dhariwal, A.~Neelakantan, P.~Shyam, G.~Sastry, A.~Askell, S.~Agarwal, A.~Herbert-Voss, G.~Krueger, T.~Henighan, R.~Child, A.~Ramesh, D.~M. Ziegler, J.~Wu, C.~Winter, C.~Hesse, M.~Chen, E.~Sigler, M.~Litwin, S.~Gray, B.~Chess, J.~Clark, C.~Berner, S.~McCandlish, A.~Radford, I.~Sutskever, and D.~Amodei, ``{Language Models are Few-Shot Learners},'' \emph{arXiv preprint arXiv:2005.14165}, 2020.

\bibitem{chen2023benchmarking}
J.~Chen, H.~Lin, X.~Han, and L.~Sun, ``{Benchmarking Large Language Models in Retrieval-Augmented Generation},'' 2023.

\bibitem{chen2023minigptv2}
J.~Chen, D.~Zhu, X.~Shen, X.~Li, Z.~Liu, P.~Zhang, R.~Krishnamoorthi, V.~Chandra, Y.~Xiong, and M.~Elhoseiny, ``{MiniGPT-v2: large language model as a unified interface for vision-language multi-task learning},'' 2023.

\bibitem{cxl}
{CXL™ Consortium}, ``{CXL 3.0 Specification},'' \url{https://computeexpresslink.org/wp-content/uploads/2023/12/CXL_3.0_white-paper_FINAL.pdf}, 2022.

\bibitem{dai2024deepseekmoe}
D.~Dai, C.~Deng, C.~Zhao, R.~X. Xu, H.~Gao, D.~Chen, J.~Li, W.~Zeng, X.~Yu, Y.~Wu, Z.~Xie, Y.~K. Li, P.~Huang, F.~Luo, C.~Ruan, Z.~Sui, and W.~Liang, ``{DeepSeekMoE: Towards Ultimate Expert Specialization in Mixture-of-Experts Language Models},'' 2024.

\bibitem{dao2023flashattention2}
T.~Dao, ``{FlashAttention-2: Faster Attention with Better Parallelism and Work Partitioning},'' \emph{arXiv preprint arXiv:2307.08691}, 2023.

\bibitem{deepseekai2024deepseekv2}
DeepSeek-AI, A.~Liu, B.~Feng, B.~Wang, B.~Wang, B.~Liu, C.~Zhao, C.~Dengr, C.~Ruan, D.~Dai, D.~Guo, D.~Yang, D.~Chen, D.~Ji, E.~Li, F.~Lin, F.~Luo, G.~Hao, G.~Chen, G.~Li, H.~Zhang, H.~Xu, H.~Yang, H.~Zhang, H.~Ding, H.~Xin, H.~Gao, H.~Li, H.~Qu, J.~L. Cai, J.~Liang, J.~Guo, J.~Ni, J.~Li, J.~Chen, J.~Yuan, J.~Qiu, J.~Song, K.~Dong, K.~Gao, K.~Guan, L.~Wang, L.~Zhang, L.~Xu, L.~Xia, L.~Zhao, L.~Zhang, M.~Li, M.~Wang, M.~Zhang, M.~Zhang, M.~Tang, M.~Li, N.~Tian, P.~Huang, P.~Wang, P.~Zhang, Q.~Zhu, Q.~Chen, Q.~Du, R.~J. Chen, R.~L. Jin, R.~Ge, R.~Pan, R.~Xu, R.~Chen, S.~S. Li, S.~Lu, S.~Zhou, S.~Chen, S.~Wu, S.~Ye, S.~Ma, S.~Wang, S.~Zhou, S.~Yu, S.~Zhou, S.~Zheng, T.~Wang, T.~Pei, T.~Yuan, T.~Sun, W.~L. Xiao, W.~Zeng, W.~An, W.~Liu, W.~Liang, W.~Gao, W.~Zhang, X.~Q. Li, X.~Jin, X.~Wang, X.~Bi, X.~Liu, X.~Wang, X.~Shen, X.~Chen, X.~Chen, X.~Nie, X.~Sun, X.~Wang, X.~Liu, X.~Xie, X.~Yu, X.~Song, X.~Zhou, X.~Yang, X.~Lu, X.~Su, Y.~Wu, Y.~K. Li, Y.~X. Wei, Y.~X. Zhu, Y.~Xu, Y.~Huang, Y.~Li, Y.~Zhao, Y.~Sun, Y.~Li,
  Y.~Wang, Y.~Zheng, Y.~Zhang, Y.~Xiong, Y.~Zhao, Y.~He, Y.~Tang, Y.~Piao, Y.~Dong, Y.~Tan, Y.~Liu, Y.~Wang, Y.~Guo, Y.~Zhu, Y.~Wang, Y.~Zou, Y.~Zha, Y.~Ma, Y.~Yan, Y.~You, Y.~Liu, Z.~Z. Ren, Z.~Ren, Z.~Sha, Z.~Fu, Z.~Huang, Z.~Zhang, Z.~Xie, Z.~Hao, Z.~Shao, Z.~Wen, Z.~Xu, Z.~Zhang, Z.~Li, Z.~Wang, Z.~Gu, Z.~Li, and Z.~Xie, ``{DeepSeek-V2: A Strong, Economical, and Efficient Mixture-of-Experts Language Model},'' 2024.

\bibitem{dettmers2024qlora}
T.~Dettmers, A.~Pagnoni, A.~Holtzman, and L.~Zettlemoyer, ``{Qlora: Efficient finetuning of quantized llms},'' \emph{Advances in Neural Information Processing Systems}, vol.~36, 2024.

\bibitem{palme}
D.~Driess, F.~Xia, M.~S.~M. Sajjadi, C.~Lynch, A.~Chowdhery, B.~Ichter, A.~Wahid, J.~Tompson, Q.~Vuong, T.~Yu, W.~Huang, Y.~Chebotar, P.~Sermanet, D.~Duckworth, S.~Levine, V.~Vanhoucke, K.~Hausman, M.~Toussaint, K.~Greff, A.~Zeng, I.~Mordatch, and P.~Florence, ``{PaLM-E: An Embodied Multimodal Language Model},'' 2023.

\bibitem{duan2024muxserve}
J.~Duan, R.~Lu, H.~Duanmu, X.~Li, X.~Zhang, D.~Lin, I.~Stoica, and H.~Zhang, ``{MuxServe: Flexible Multiplexing for Efficient Multiple LLM Serving},'' 2024.

\bibitem{frantar2023sparsegpt}
E.~Frantar and D.~Alistarh, ``{SparseGPT: Massive Language Models Can Be Accurately Pruned in One-Shot},'' in \emph{International Conference on Machine Learning}.\hskip 1em plus 0.5em minus 0.4em\relax PMLR, 2023, pp. 10\,323--10\,337.

\bibitem{frantar-gptq}
E.~Frantar, S.~Ashkboos, T.~Hoefler, and D.~Alistarh, ``{GPTQ: Accurate Post-training Compression for Generative Pretrained Transformers},'' \emph{arXiv preprint arXiv:2210.17323}, 2022.

\bibitem{gemmini-dac}
H.~Genc, S.~Kim, A.~Amid, A.~Haj-Ali, V.~Iyer, P.~Prakash, J.~Zhao, D.~Grubb, H.~Liew, H.~Mao, A.~Ou, C.~Schmidt, S.~Steffl, J.~Wright, I.~Stoica, J.~Ragan-Kelley, K.~Asanovic, B.~Nikolic, and Y.~S. Shao, ``{Gemmini: Enabling Systematic Deep-Learning Architecture Evaluation via Full-Stack Integration},'' in \emph{Proceedings of the 58th Annual Design Automation Conference (DAC)}, 2021.

\bibitem{genesys}
S.~Ghodrati, S.~Kinzer, H.~Xu, R.~Mahapatra, B.~H. Ahn, D.~K. Wang, L.~Karthikeyan, A.~Yazdanbakhsh, J.~Park, N.~S. Kim, and H.~Esmaeilzadeh, ``{Tandem Processor: Grappling with Emerging Operators in Neural Networks},'' in \emph{ASPLOS}, 2024.

\bibitem{onnxim}
\BIBentryALTinterwordspacing
H.~Ham, W.~Yang, Y.~Shin, O.~Woo, G.~Heo, S.~Lee, J.~Park, and G.~Kim, ``{ONNXim: A Fast, Cycle-level Multi-core NPU Simulator},'' 2024. [Online]. Available: \url{https://arxiv.org/abs/2406.08051}
\BIBentrySTDinterwordspacing

\bibitem{ham2020a3}
T.~Ham, S.~Jung, S.~Kim, Y.~H. Oh, Y.~Park, Y.~Song, J.~Park, S.~Lee, K.~Park, J.~W. Lee, and D.~Jeong, ``{A3: Accelerating Attention Mechanisms in Neural Networks with Approximation},'' in \emph{HPCA}, 2020.

\bibitem{ham2021elsa}
T.~J. Ham, Y.~Lee, S.~H. Seo, S.~Kim, H.~Choi, S.~J. Jung, and J.~W. Lee, ``{ELSA: Hardware-Software Co-design for Efficient, Lightweight Self-Attention Mechanism in Neural Networks},'' in \emph{ISCA}, 2021.

\bibitem{neupims}
G.~Heo, S.~Lee, J.~Cho, H.~Choi, S.~Lee, H.~Ham, G.~Kim, D.~Mahajan, and J.~Park, ``{NeuPIMs: NPU-PIM Heterogeneous Acceleration for Batched LLM Inferencing},'' in \emph{ASPLOS}, 2024.

\bibitem{hong2022dfx}
S.~Hong, S.~Moon, J.~Kim, S.~Lee, M.~Kim, D.~Lee, and J.-Y. Kim, ``{DFX: A Low-latency Multi-FPGA Appliance for Accelerating Transformer-based Text Generation},'' in \emph{MICRO}, 2022.

\bibitem{hooper2024kvquant}
C.~Hooper, S.~Kim, H.~Mohammadzadeh, M.~W. Mahoney, Y.~S. Shao, K.~Keutzer, and A.~Gholami, ``{KVQuant: Towards 10 Million Context Length LLM Inference with KV Cache Quantization},'' \emph{arXiv preprint arXiv:2401.18079}, 2024.

\bibitem{hu2021lora}
E.~J. Hu, Y.~Shen, P.~Wallis, Z.~Allen-Zhu, Y.~Li, S.~Wang, L.~Wang, and W.~Chen, ``{Lora: Low-rank adaptation of large language models},'' \emph{arXiv preprint arXiv:2106.09685}, 2021.

\bibitem{hu2023llm}
Z.~Hu, L.~Wang, Y.~Lan, W.~Xu, E.-P. Lim, L.~Bing, X.~Xu, S.~Poria, and R.~K.-W. Lee, ``{LLM-adapters: An adapter family for parameter-efficient fine-tuning of large language models},'' \emph{arXiv preprint arXiv:2304.01933}, 2023.

\bibitem{mnpusim}
S.~Hwang, S.~Lee, J.~Kim, H.~Kim, and J.~Huh, ``{mNPUsim: Evaluating the Effect of Sharing Resources in Multi-core NPUs},'' in \emph{2023 IEEE International Symposium on Workload Characterization (IISWC)}.\hskip 1em plus 0.5em minus 0.4em\relax IEEE, 2023, pp. 167--179.

\bibitem{upmem-hpca24}
B.~Hyun, T.~Kim, D.~Lee, and M.~Rhu, ``{Pathfinding Future PIM Architectures by Demystifying a Commercial PIM Technology},'' in \emph{HPCA}, 2024.

\bibitem{smart-infinity}
H.~Jang, J.~Song, J.~Jung, J.~Park, Y.~Kim, and J.~Lee, ``{Smart-Infinity: Fast Large Language Model Training using Near-Storage Processing on a Real System},'' in \emph{HPCA}, 2024.

\bibitem{jiang2023active}
Z.~Jiang, F.~F. Xu, L.~Gao, Z.~Sun, Q.~Liu, J.~Dwivedi-Yu, Y.~Yang, J.~Callan, and G.~Neubig, ``{Active Retrieval Augmented Generation},'' 2023.

\bibitem{sait-pimsim}
S.~Kang, S.~Cha, S.~Seo, and J.~Kim, ``{PIMSimulator},'' \url{https://github.com/SAITPublic/PIMSimulator}, 2022.

\bibitem{polymath}
S.~Kinzer, J.~K. Kim, S.~Ghodrati, B.~Yatham, A.~Althoff, D.~Mahajan, S.~Lerner, and H.~Esmaeilzadeh, ``{A Computational Stack for Cross-Domain Acceleration},'' in \emph{HPCA}, 2021.

\bibitem{vllm}
W.~Kwon, Z.~Li, S.~Zhuang, Y.~Sheng, L.~Zheng, C.~H. Yu, J.~E. Gonzalez, H.~Zhang, and I.~Stoica, ``{Efficient Memory Management for Large Language Model Serving with PagedAttention},'' \emph{arXiv preprint arXiv:2309.06180}, 2023.

\bibitem{hbm-pim}
S.~Lee, S.-h. Kang, J.~Lee, H.~Kim, E.~Lee, S.~Seo, H.~Yoon, S.~Lee, K.~Lim, H.~Shin, J.~Kim, O.~Seongil, A.~Iyer, D.~Wang, K.~Sohn, and N.~S. Kim, ``{Hardware Architecture and Software Stack for PIM Based on Commercial DRAM Technology : Industrial Product},'' in \emph{ISCA}, 2021.

\bibitem{lewis2020retrieval}
P.~Lewis, E.~Perez, A.~Piktus, F.~Petroni, V.~Karpukhin, N.~Goyal, H.~K{\"u}ttler, M.~Lewis, W.-t. Yih, T.~Rockt{\"a}schel \emph{et~al.}, ``{Retrieval-Augmented Generation for Knowledge-Intensive NLP Tasks},'' \emph{Advances in Neural Information Processing Systems}, vol.~33, pp. 9459--9474, 2020.

\bibitem{moe-lina}
\BIBentryALTinterwordspacing
J.~Li, Y.~Jiang, Y.~Zhu, C.~Wang, and H.~Xu, ``{Accelerating Distributed {MoE} Training and Inference with Lina},'' in \emph{2023 USENIX Annual Technical Conference (USENIX ATC 23)}.\hskip 1em plus 0.5em minus 0.4em\relax Boston, MA: USENIX Association, Jul. 2023, pp. 945--959. [Online]. Available: \url{https://www.usenix.org/conference/atc23/presentation/li-jiamin}
\BIBentrySTDinterwordspacing

\bibitem{liu2024llavanext}
\BIBentryALTinterwordspacing
H.~Liu, C.~Li, Y.~Li, B.~Li, Y.~Zhang, S.~Shen, and Y.~J. Lee, ``{LLaVA-NeXT: Improved reasoning, OCR, and world knowledge},'' January 2024. [Online]. Available: \url{https://llava-vl.github.io/blog/2024-01-30-llava-next/}
\BIBentrySTDinterwordspacing

\bibitem{lu2020ha}
S.~Lu, M.~Wang, S.~Liang, J.~Lin, and Z.~Wang, ``{Hardware Accelerator for Multi-Head Attention and Position-Wise Feed-Forward in the Transformer},'' in \emph{2020 IEEE 33rd International System-on-Chip Conference (SOCC)}, 2020, pp. 84--89.

\bibitem{ma2024era}
S.~Ma, H.~Wang, L.~Ma, L.~Wang, W.~Wang, S.~Huang, L.~Dong, R.~Wang, J.~Xue, and F.~Wei, ``{The Era of 1-bit LLMs: All Large Language Models are in 1.58 Bits},'' \emph{arXiv preprint arXiv:2402.17764}, 2024.

\bibitem{ma2023llm}
X.~Ma, G.~Fang, and X.~Wang, ``{LLM-pruner: On the structural pruning of large language models},'' \emph{Advances in neural information processing systems}, vol.~36, pp. 21\,702--21\,720, 2023.

\bibitem{miao2024spotserve}
X.~Miao, C.~Shi, J.~Duan, X.~Xi, D.~Lin, B.~Cui, and Z.~Jia, ``{SpotServe: Serving Generative Large Language Models on Preemptible Instances},'' in \emph{ASPLOS}, 2024.

\bibitem{onnx}
F.~R. Microsoft, ``{ONNX: an open format to represent deep learning models.}'' \url{http://onnx.ai/}, 2017.

\bibitem{dist-gem5}
A.~Mohammad, U.~Darbaz, G.~Dozsa, S.~Diestelhorst, D.~Kim, and N.~S. Kim, ``{dist-gem5: Distributed simulation of computer clusters},'' in \emph{2017 IEEE International Symposium on Performance Analysis of Systems and Software (ISPASS)}, 2017, pp. 153--162.

\bibitem{naveed2024comprehensive}
H.~Naveed, A.~U. Khan, S.~Qiu, M.~Saqib, S.~Anwar, M.~Usman, N.~Akhtar, N.~Barnes, and A.~Mian, ``{A Comprehensive Overview of Large Language Models},'' \emph{arXiv preprint arXiv:2307.06435}, 2024.

\bibitem{nvidia-h100}
NVIDIA, ``{NVIDIA H100 Tensor Core GPU Architecture},'' \url{https://resources.nvidia.com/en-us-data-center-overview-mc/en-us-data-center-overview/gtc22-whitepaper-hopper?pflpid=17841&lb-mode=preview}, 2023.

\bibitem{exegpt}
H.~Oh, K.~Kim, J.~Kim, S.~Kim, J.~Lee, D.-s. Chang, and J.~Seo, ``{ExeGPT: Constraint-Aware Resource Scheduling for LLM Inference},'' in \emph{ASPLOS}, 2024.

\bibitem{chatgpt}
OpenAI, ``{ChatGPT},'' \url{ https://chatgpt.com/blog/chatgpt}, 2024.

\bibitem{attacc}
J.~Park, J.~Choi, K.~Kyung, M.~J. Kim, Y.~Kwon, N.~S. Kim, and J.~H. Ahn, ``{AttAcc! Unleashing the Power of PIM for Batched Transformer-based Generative Model Inference},'' in \emph{ASPLOS}, 2024.

\bibitem{park2024lpddr}
S.-S. Park, K.~Kim, J.~So, J.~Jung, J.~Lee, K.~Woo, N.~Kim, Y.~Lee, H.~Kim, Y.~Kwon, J.~Kim, J.~Lee, Y.~Cho, Y.~Tai, J.~Cho, H.~Song, J.~H. Ahn, and N.~S. Kim, ``{An LPDDR-based CXL-PNM Platform for TCO-efficient Inference of Transformer-based Large Language Models},'' in \emph{HPCA}.\hskip 1em plus 0.5em minus 0.4em\relax IEEE, 2024, pp. 970--982.

\bibitem{torch}
A.~Paszke, S.~Gross, F.~Massa, A.~Lerer, J.~Bradbury, G.~Chanan, T.~Killeen, Z.~Lin, N.~Gimelshein, L.~Antiga, A.~Desmaison, A.~Köpf, E.~Yang, Z.~DeVito, M.~Raison, A.~Tejani, S.~Chilamkurthy, B.~Steiner, L.~Fang, J.~Bai, and S.~Chintala, ``{PyTorch: An Imperative Style, High-Performance Deep Learning Library},'' \emph{arXiv preprint arXiv:1912.01703}, 2019.

\bibitem{splitwise}
\BIBentryALTinterwordspacing
P.~Patel, E.~Choukse, C.~Zhang, A.~Shah, I.~Goiri, S.~Maleki, and R.~Bianchini, ``{Splitwise: Efficient Generative LLM Inference Using Phase Splitting},'' in \emph{2024 ACM/IEEE 51st Annual International Symposium on Computer Architecture (ISCA)}.\hskip 1em plus 0.5em minus 0.4em\relax Los Alamitos, CA, USA: IEEE Computer Society, jul 2024, pp. 118--132. [Online]. Available: \url{https://doi.ieeecomputersociety.org/10.1109/ISCA59077.2024.00019}
\BIBentrySTDinterwordspacing

\bibitem{patel2023splitwise}
P.~Patel, E.~Choukse, C.~Zhang, Íñigo Goiri, A.~Shah, S.~Maleki, and R.~Bianchini, ``{Splitwise: Efficient generative LLM inference using phase splitting},'' \emph{arXiv preprint arXiv:2311.18677}, 2023.

\bibitem{peng2023kosmos2}
Z.~Peng, W.~Wang, L.~Dong, Y.~Hao, S.~Huang, S.~Ma, and F.~Wei, ``{Kosmos-2: Grounding Multimodal Large Language Models to the World},'' 2023.

\bibitem{clip}
A.~Radford, J.~W. Kim, C.~Hallacy, A.~Ramesh, G.~Goh, S.~Agarwal, G.~Sastry, A.~Askell, P.~Mishkin, J.~Clark, G.~Krueger, and I.~Sutskever, ``{Learning Transferable Visual Models From Natural Language Supervision},'' 2021.

\bibitem{ram2023context}
O.~Ram, Y.~Levine, I.~Dalmedigos, D.~Muhlgay, A.~Shashua, K.~Leyton-Brown, and Y.~Shoham, ``{In-context retrieval-augmented language models},'' \emph{Transactions of the Association for Computational Linguistics}, vol.~11, pp. 1316--1331, 2023.

\bibitem{dts}
W.~J. Robinson~M., F.~Esposito, and M.~A. Zuluaga, ``{DTS: A Simulator to Estimate the Training Time of Distributed Deep Neural Networks},'' in \emph{2022 30th International Symposium on Modeling, Analysis, and Simulation of Computer and Telecommunication Systems (MASCOTS)}, 2022, pp. 17--24.

\bibitem{simtool}
\BIBentryALTinterwordspacing
A.~F. Rodrigues, K.~S. Hemmert, B.~W. Barrett, C.~Kersey, R.~Oldfield, M.~Weston, R.~Risen, J.~Cook, P.~Rosenfeld, E.~Cooper-Balis, and B.~Jacob, ``{The structural simulation toolkit},'' \emph{SIGMETRICS Perform. Eval. Rev.}, vol.~38, no.~4, p. 37–42, mar 2011. [Online]. Available: \url{https://doi.org/10.1145/1964218.1964225}
\BIBentrySTDinterwordspacing

\bibitem{samajdar2019scalesim}
A.~Samajdar, Y.~Zhu, P.~Whatmough, M.~Mattina, and T.~Krishna, ``{SCALE-Sim: Systolic CNN Accelerator Simulator},'' 2019.

\bibitem{ianus}
M.~Seo, X.~T. Nguyen, S.~J. Hwang, Y.~Kwon, G.~Kim, C.~Park, I.~Kim, J.~Park, J.~Kim, W.~Shin, J.~Won, H.~Choi, K.~Kim, D.~Kwon, C.~Jeong, S.~Lee, Y.~Choi, W.~Byun, S.~Baek, H.-J. Lee, and J.~Kim, ``{IANUS: Integrated Accelerator based on NPU-PIM Unified Memory System},'' in \emph{ASPLOS}, 2024.

\bibitem{gem5_aladdin}
Y.~S. Shao, S.~L. Xi, V.~Srinivasan, G.-Y. Wei, and D.~Brooks, ``{Co-designing accelerators and SoC interfaces using gem5-Aladdin},'' in \emph{MICRO}, 2016.

\bibitem{shao2023enhancing}
Z.~Shao, Y.~Gong, Y.~Shen, M.~Huang, N.~Duan, and W.~Chen, ``{Enhancing Retrieval-Augmented Large Language Models with Iterative Retrieval-Generation Synergy},'' 2023.

\bibitem{sharegpt}
{ShareGPT Team}, ``{ShareGPT},'' \url{https://sharegpt.com}, 2023.

\bibitem{shen2023mars}
G.~Shen, J.~Zhao, Z.~Wang, Z.~Lin, W.~Ding, C.~Wu, Q.~Chen, and M.~Guo, ``{MARS: Exploiting Multi-Level Parallelism for DNN Workloads on Adaptive Multi-Accelerator Systems},'' \emph{arXiv preprint arXiv:2307.12234}, 2023.

\bibitem{megatron}
M.~Shoeybi, M.~Patwary, R.~Puri, P.~LeGresley, J.~Casper, and B.~Catanzaro, ``{Megatron-LM: Training Multi-Billion Parameter Language Models Using Model Parallelism},'' \emph{arXiv preprint arXiv:1909.08053}, 2020.

\bibitem{chakra}
S.~Sridharan, T.~Heo, L.~Feng, Z.~Wang, M.~Bergeron, W.~Fu, S.~Zheng, B.~Coutinho, S.~Rashidi, C.~Man, and T.~Krishna, ``{Chakra: Advancing Performance Benchmarking and Co-design using Standardized Execution Traces},'' \emph{arXiv preprint arXiv:2305.14516}, 2023.

\bibitem{sun2023simple}
M.~Sun, Z.~Liu, A.~Bair, and J.~Z. Kolter, ``{A simple and effective pruning approach for large language models},'' \emph{arXiv preprint arXiv:2306.11695}, 2023.

\bibitem{edgebert}
T.~Tambe, C.~Hooper, L.~Pentecost, T.~Jia, E.-Y. Yang, M.~Donato, V.~Sanh, P.~Whatmough, A.~M. Rush, D.~Brooks, and G.-Y. Wei, ``{EdgeBERT: Sentence-Level Energy Optimizations for Latency-Aware Multi-Task NLP Inference},'' in \emph{MICRO}, 2021.

\bibitem{COSSIM}
\BIBentryALTinterwordspacing
N.~Tampouratzis, I.~Papaefstathiou, A.~Nikitakis, A.~Brokalakis, S.~Andrianakis, A.~Dollas, M.~Marcon, and E.~Plebani, ``{A Novel, Highly Integrated Simulator for Parallel and Distributed Systems},'' \emph{ACM Trans. Archit. Code Optim.}, vol.~17, no.~1, mar 2020. [Online]. Available: \url{https://doi.org/10.1145/3378934}
\BIBentrySTDinterwordspacing

\bibitem{alpaca}
R.~Taori, I.~Gulrajani, T.~Zhang, Y.~Dubois, X.~Li, C.~Guestrin, P.~Liang, and T.~B. Hashimoto, ``{Stanford Alpaca: An Instruction-following LLaMA model},'' \url{https://github.com/tatsu-lab/stanford_alpaca}, 2023.

\bibitem{chameleonteam2024chameleon}
C.~Team, ``{Chameleon: Mixed-Modal Early-Fusion Foundation Models},'' 2024.

\bibitem{llama}
H.~Touvron, T.~Lavril, G.~Izacard, X.~Martinet, M.-A. Lachaux, T.~Lacroix, B.~Rozière, N.~Goyal, E.~Hambro, F.~Azhar, A.~Rodriguez, A.~Joulin, E.~Grave, and G.~Lample, ``{LLaMA: Open and Efficient Foundation Language Models},'' \emph{arXiv preprint arXiv:2302.13971}, 2023.

\bibitem{vaswani2023attention}
A.~Vaswani, N.~Shazeer, N.~Parmar, J.~Uszkoreit, L.~Jones, A.~N. Gomez, L.~u. Kaiser, and I.~Polosukhin, ``{Attention is All you Need},'' in \emph{Advances in Neural Information Processing Systems}, 2017.

\bibitem{spatten}
H.~Wang, Z.~Zhang, and S.~Han, ``{SpAtten: Efficient Sparse Attention Architecture with Cascade Token and Head Pruning},'' in \emph{HPCA}, 2021.

\bibitem{astrasim}
W.~Won, T.~Heo, S.~Rashidi, S.~Sridharan, S.~Srinivasan, and T.~Krishna, ``{ASTRA-sim2.0: Modeling Hierarchical Networks and Disaggregated Systems for Large-model Training at Scale},'' in \emph{2023 IEEE International Symposium on Performance Analysis of Systems and Software (ISPASS)}, 2023.

\bibitem{wu2023fast}
B.~Wu, Y.~Zhong, Z.~Zhang, G.~Huang, X.~Liu, and X.~Jin, ``{Fast Distributed Inference Serving for Large Language Models},'' 2023.

\bibitem{nextgpt}
S.~Wu, H.~Fei, L.~Qu, W.~Ji, and T.-S. Chua, ``{NExT-GPT: Any-to-Any Multimodal LLM},'' 2023.

\bibitem{grok}
xAI, ``{Grok-1},'' \url{ https://github.com/xai-org/grok-1}, 2024.

\bibitem{xiao2023smoothquant}
G.~Xiao, J.~Lin, M.~Seznec, H.~Wu, J.~Demouth, and S.~Han, ``{SmoothQuant: Accurate and Efficient Post-Training Quantization for Large Language Models},'' in \emph{Proceedings of the 40th International Conference on Machine Learning}, 2023.

\bibitem{xu2024retrieval}
P.~Xu, W.~Ping, X.~Wu, L.~McAfee, C.~Zhu, Z.~Liu, S.~Subramanian, E.~Bakhturina, M.~Shoeybi, and B.~Catanzaro, ``{Retrieval meets Long Context Large Language Models},'' 2024.

\bibitem{yao2022zeroquant}
Z.~Yao, R.~Yazdani~Aminabadi, M.~Zhang, X.~Wu, C.~Li, and Y.~He, ``{Zeroquant: Efficient and affordable post-training quantization for large-scale transformers},'' \emph{Advances in Neural Information Processing Systems}, vol.~35, pp. 27\,168--27\,183, 2022.

\bibitem{orca}
G.-I. Yu, J.~S. Jeong, G.-W. Kim, S.~Kim, and B.-G. Chun, ``{Orca: A Distributed Serving System for {Transformer-Based} Generative Models},'' in \emph{16th USENIX Symposium on Operating Systems Design and Implementation (OSDI 22)}.\hskip 1em plus 0.5em minus 0.4em\relax Carlsbad, CA: USENIX Association, Jul. 2022, pp. 521--538.

\bibitem{gobo}
A.~Zadeh, I.~Edo, O.~Awad, and A.~Moshovos, ``{GOBO: Quantizing Attention-Based NLP Models for Low Latency and Energy Efficient Inference},'' in \emph{MICRO}, 2020.

\bibitem{zhao2023survey}
W.~X. Zhao, K.~Zhou, J.~Li, T.~Tang, X.~Wang, Y.~Hou, Y.~Min, B.~Zhang, J.~Zhang, Z.~Dong, Y.~Du, C.~Yang, Y.~Chen, Z.~Chen, J.~Jiang, R.~Ren, Y.~Li, X.~Tang, Z.~Liu, P.~Liu, J.-Y. Nie, and J.-R. Wen, ``{A Survey of Large Language Models},'' 2023.

\bibitem{zhao2024alisa}
Y.~Zhao, D.~Wu, and J.~Wang, ``{ALISA: Accelerating Large Language Model Inference via Sparsity-Aware KV Caching},'' 2024.

\bibitem{distserve}
\BIBentryALTinterwordspacing
Y.~Zhong, S.~Liu, J.~Chen, J.~Hu, Y.~Zhu, X.~Liu, X.~Jin, and H.~Zhang, ``{{DistServe}: Disaggregating Prefill and Decoding for Goodput-optimized Large Language Model Serving},'' in \emph{18th USENIX Symposium on Operating Systems Design and Implementation (OSDI 24)}.\hskip 1em plus 0.5em minus 0.4em\relax Santa Clara, CA: USENIX Association, Jul. 2024, pp. 193--210. [Online]. Available: \url{https://www.usenix.org/conference/osdi24/presentation/zhong-yinmin}
\BIBentrySTDinterwordspacing

\bibitem{transpim}
M.~Zhou, W.~Xu, J.~Kang, and T.~Rosing, ``{TransPIM: A Memory-based Acceleration via Software-Hardware Co-Design for Transformer},'' in \emph{HPCA}, 2022.

\bibitem{minigpt4}
D.~Zhu, J.~Chen, X.~Shen, X.~Li, and M.~Elhoseiny, ``{MiniGPT-4: Enhancing Vision-Language Understanding with Advanced Large Language Models},'' 2023.

\end{thebibliography}

%
%
%
%
%






\clearpage

\appendix
\section{Artifact Appendix}

\subsection{Abstract}
\simname is a fast and accurate hardware-software co-simulation infrastructure for LLM inference serving systems written with C++ and Python.
\simname receives several system configurations and request traces from the user, calculates the cycles and throughput of the system composed of various accelerators, and measures the inference latency for each request.
\subsection{Artifact Check-list (Meta-information)}

{\small
\begin{itemize}
  \item {{\bf Compilation:} g++ v7.5.0}
  \item {{\bf Run-time environment:} Ubuntu 18.04 Kernel v4.15.0}
  \item {{\bf Hardware:} x86-64}
  \item {{\bf Output:} standard output, \textit{TSV} files}
  \item {{\bf How much disk space required (approximately)?:} Artifact evaluation requires up to 30GB of disk space, but depending on the models and datasets, it may require 1GB $\sim$ 400GB of disk space.}
  \item {{\bf How much time is needed to prepare workflow (approximately)?:} 5 minutes}
  \item {{\bf How much time is needed to complete experiments (approximately)?:} Artifact evaluation takes approximately 12 hours, but depending on the models and datasets, it may take 30 seconds $\sim$ 24 hours.}
  \item {{\bf Publicly available?:} Yes}
  \item {{\bf Code licenses (if publicly available)?:} Creative Commons Attribution 4.0 International, MIT License}
  \item {{\bf Workflow framework used?:} No}
  \item {{\bf Archived (provide DOI)?:} Yes (10.5281/zenodo.12803583)}
\end{itemize}
}

\subsection{Description}

\subsubsection{How to access}
{
\begin{itemize}
    \item {Zenodo: \simname is published on Zenodo: 

    \bluetext{\url{https://doi.org/10.5281/zenodo.12803583}}}
    \item {GitHub: \simname is available on GitHub: 
    
    \bluetext{\url{https://github.com/casys-kaist/llmservingsim}}}
\end{itemize}
}
\subsubsection{Hardware dependencies}
\simname requires an x86-64 architecture, and the simulation time may be affected by hardware differences.
For similar simulation time results, we recommend using the hardware specified in Section~\ref{sec:methodology}.
\subsubsection{Software dependencies}
\simname has been tested on Ubuntu 18.04 with Python 3.9 and requires gcc and g++ versions 7.5.0 or higher.
Additionally, it requires the software prerequisites of ASTRA-Sim~\cite{astrasim}, Chakra~\cite{chakra}, and Polymath~\cite{polymath}. 
To meet these software prerequisites, we use the latest version of Conda. 
We provide the instructions for Conda installation in Appendix~\ref{sec:installation} and \textit{README} file.
Also, it can be downloaded individually from the following link:~\bluetext{\url{https://repo.anaconda.com/archive/}}.
\subsubsection{Data sets}
We use ShareGPT~\cite{sharegpt} and Alpaca~\cite{alpaca} datasets to generate arbitrary request trace.
\subsubsection{Models}
We use GPT3~\cite{gpt3} and LLaMA~\cite{llama} with model size of 7B to 175B for our evaluation.
Their model architecture follows the decoder-based transformer model.
%
\subsection{Installation}
\label{sec:installation}
{
\begin{itemize}
\item {Clone the \simname repository.
\begin{lstlisting}[language=bash]
$ git clone --recurse-submodules https://github.com/casys-kaist/LLMServingSim.git
$ cd LLMServingSim
\end{lstlisting}
}
\item {Conda install (optional).
\begin{lstlisting}[language=bash]
$ curl -O https://repo.anaconda.com/archive/Anaconda3-2024.06-1-Linux-x86_64.sh
$ bash Anaconda3-2024.06-1-Linux-x86_64.sh
\end{lstlisting}
}
\item {Install dependencies.
\begin{lstlisting}[language=bash]
$ conda env create -p ./env -f ./environment.yml
$ conda activate ./env
\end{lstlisting}
}
\item {Build submodules.
\begin{lstlisting}[language=bash]
$ cd astra-sim
$ ./build/astra_analytical/build.sh
$ cd extern/graph_frontend/chakra
$ pip install .
$ cd ../../../../execution_engine/polymath
$ pip install .
$ cd ../..
\end{lstlisting}
}
\end{itemize}
}

\subsection{Experiment Workflow}
The workflow of \simname is well illustrated in Section~\ref{sec:infrastructure}, particularly in Figure~\ref{fig:llmsim}. 
To explain this in detail, the simulator first receives the system configuration and request trace from the user.
Then, the scheduler selects requests that can be batched in each iteration using the KV cache information and creates a simulation plan that maps operators to each execution engine.
Each engine simulates the operators using this simulation plan, and the results are combined into a single trace through operator scheduling.
This trace goes through a graph converter to become an execution graph~\cite{chakra}, which is then simulated at the system level by ASTRA-Sim~\cite{astrasim}.
Finally, the execution results are returned to the scheduler, which uses them to proceed to the next iteration.
In this process, the simulator calculates the system's throughput and latency.

\subsection{Evaluation and Expected Results}
The evaluation conducted in this paper can be categorized into five parts,  as described in Section \ref{sec:evaluation}.
To facilitate the execution of these five evaluations, we have created five separate scripts, stored in the \textit{evaluation} folder, each for running an individual experiment.
We also provide a script (\textit{evaluation\_all.sh}) to run all of them at once.
{
\begin{itemize}
\item {Move to \textit{evaluation} folder.
\begin{lstlisting}[language=bash]
$ cd evaluation
\end{lstlisting}
}
\item {Run each evaluation one by one.
\begin{lstlisting}[language=bash]
$ ./evaluation1.sh
$ ./evaluation2.sh
...
$ ./evaluation5.sh
\end{lstlisting}
}
\item {Run all evaluation at once.
\begin{lstlisting}[language=bash]
$ ./evaluation_all.sh
\end{lstlisting}
}
\end{itemize}
}
The results of each script are stored in their respective evaluation folders.
Each command within a script generates three files, including (1)  \textit{text} file containing the redirected standard output and (2) two \textit{TSV} files storing throughput and simulation time.

To facilitate verification of the results used in the paper, we provide an \textit{Excel} file (\textit{evaluation.xlsx}) in the \textit{evaluation} folder. 
The \textit{Excel} file contains the numbers and figures from the paper, along with instructions to manipulate the raw data. 
For more information, please refer to Section~\ref{sec:evaluation} and the \textit{README} files in each folder.
Figure~\ref{fig:dirtree} illustrates the directory tree of \simname with the evaluation scripts, their outputs, \textit{Excel} file, and \textit{README} files.
\begin{figure}
\centering
\framebox[0.9\linewidth]{%
\begin{minipage}{0.8\linewidth}
\dirtree{%
    .1 \simname.
    .2 evaluation.
    .3 evaluation1.
    .4 gpt7b.txt.
    .4 gpt7b-throughput.tsv.
    .4 gpt7b-simulation-time.tsv.
    .4 \vdots.
    .3 \vdots.
    .3 evaluation1.sh.
    .3 \vdots.
    .3 evaluation{\_}all.sh.
    .3 evaluation.xlsx.
    .3 README.md.
    .2 \vdots.
    .2 README.md.
}
\end{minipage}
}
\caption{Directory tree of \simname.}
\label{fig:dirtree}
\end{figure}

\subsection{Experiment Customization}
\subsubsection{Input configurations}
\simname takes configuration files for hardware and network as input.
{
\begin{itemize}
\item {\textit{NPU config}: a \textit{json} file that contains configurations of the NPU. It is located at \textit{execution{\_}engine/codelets{\_}src/ codelets/examples/genesys/configs/} folder.}
\item {\textit{network config}: a \textit{json} file that contains configurations of the system network topology. It is located at \textit{astra-sim/inputs/network/analytical/} folder.}
\end{itemize}
}
\subsubsection{Input dataset}
\simname takes LLM inference request datasets with various request patterns as input.
{
\begin{itemize}
\item {\textit{dataset}: a \textit{TSV} file that contains the input token length, output token length, and arrival time. It is located at \textit{astra-sim/dataset/} folder.}
\end{itemize}
}
\subsubsection{Input parameters}
\simname has a total of 16 parameters for various simulation configurations. \textit{README} file provides usage instructions and examples.
{
\begin{itemize}
\item {\textit{model{\_}name}: Name of the LLM model. Default value is `gpt2'.}
\item {\textit{npu{\_}num}: Number of NPUs in the system. Default value is 16.}
\item {\textit{max{\_}batch}: Maximum batch size. Default value is 0, which indicates no limit.}
\item {\textit{batch{\_}delay}: Delay of batching. Default value is 0.}
\item {\textit{scheduling}: The method of scheduling. Default value is `orca', which refers to the iteration-level scheduling technique proposed in Orca~\cite{orca}.}
\item {\textit{parallel}: The method of parallelism. There are three methods: `pipeline', `tensor', and `hybrid'. Default value is `hybrid'.}
\item {\textit{npu{\_}group}: Number of NPU groups used in hybrid parallelism. Default value is 1.}
\item {\textit{npu{\_}mem}: Local memory size of the NPU in GB. Default value is 40.}
\item {\textit{kv{\_}manage}: The method of KV cache management. Default value is `vllm', which refers to the paged attention technique proposed in vLLM~\cite{vllm}.}
\item {\textit{pim{\_}type}: The method of using PIM. There are three methods: `none', `local', and `pool'. Default value is `none', which indicates no use of PIM.}
\item {\textit{sub{\_}batch}: The method of scheduling when using PIM. It is a flag that turns on for the sub-batch interleaving technique proposed in NeuPIMs~\cite{neupims}.}
\item {\textit{dataset}: The path of the dataset.}
\item {\textit{network}: The path of the network configuration file.}
\item {\textit{output}: The path of the output \textit{TSV} files.}
\item {\textit{gen}: The flag that indicates the initiation phase should be skipped when enabled.}
\item {\textit{fast{\_}run}: The flag that turns on the model compilation bypassing that facilitates the fast reproduction of evaluation 1 (simulation time: 32 hours $\rightarrow$ 20 minutes).}
\end{itemize}
}
\subsubsection{Result files}
\simname provides three outputs.
{
\begin{itemize}
\item {\textit{standard output}: It shows which requests are being processed in each iteration of the simulator and displays the measured throughput at regular intervals. Additionally, it provides a summary of throughput and simulation time at the end.}
\item {\textit{\{output{\_}filename\}-throughput.tsv}: The file stores prompt and generation throughput at regular intervals.}
\item {\textit{\{output{\_}filename\}-simulation-time.tsv}: The file stores each simulation component's simulation time in milliseconds.}
\end{itemize}
}
%
\subsection{Notes}

More information can be found in the \textit{README} file of each directory.

\balance






\end{document}